\documentclass[
  reprint,       
  aps,            
  prb,            
  superscriptaddress,
  floatfix
]{revtex4-2}

\usepackage{amsmath,amssymb,amsthm}
\usepackage{graphicx}
\usepackage{bm}
\usepackage{physics}
\usepackage{xurl}
\usepackage{braket}
\usepackage{amsfonts}
\usepackage{dsfont}
\usepackage[dvipsnames]{xcolor}
\usepackage{appendix}
\usepackage{longtable}
\usepackage{soul}
\usepackage{float}
\usepackage{algorithm}
\usepackage{algpseudocode}
\usepackage{microtype}
\usepackage[breaklinks=true]{hyperref}

\newcommand{\beq}{\begin{equation}}
\newcommand{\eeq}{\end{equation}}
\newcommand{\beqa}{\begin{eqnarray}}
\newcommand{\eeqa}{\end{eqnarray}}

\theoremstyle{definition}
\newtheorem{definition}{Definition}
\newtheorem{example}{Example}
\theoremstyle{remark}
\newtheorem{remark}{Remark}
\theoremstyle{plain}

\algrenewcommand\algorithmicrequire{\textbf{Input:}}
\algrenewcommand\algorithmicensure{\textbf{Output:}}

\begin{document}

\title{The Kitaev model on finite hyperbolic PBC clusters}
\author{Alberto Zorzato}
\affiliation{Institute for Theoretical Physics, University of Amsterdam,
Science Park 904, 1098 XH Amsterdam, The Netherlands
\\QuSoft, Science Park 123, 1098 XG Amsterdam, The Netherlands}

\author{Ji\v{r}\'{i} Min\'{a}\v{r}}
\affiliation{Institute for Theoretical Physics, University of Amsterdam,
Science Park 904, 1098 XH Amsterdam, The Netherlands
\\QuSoft, Science Park 123, 1098 XG Amsterdam, The Netherlands}

\date{\today}
\begin{abstract}
We construct and solve the three-colored Kitaev model on finite hyperbolic lattices with periodic boundary conditions.
We describe a self-contained procedure which uses finite quotients of orientation-preserving triangle groups to generate regular maps on compact genus-$g$ surfaces. From these, one can extract their lattice connectivity, plaquette structure, homology cycles and Wilson sectors while implementing consistent boundary conditions. 
On the example of $\{8,3\}$ and $\{12,3\}$ lattices, we enumerate a selection of quotients with up to 72 vertices, which constitute the discrete steps in a finite-size scaling analysis. 
For the isotropic point, corresponding to the spin-liquid phase in the thermodynamic limit, we identify the ground state superselection sectors, quantify their deviation from the homogeneous $\pi$ flux sector and show how the selected flux background depends on the quotient geometry. 
We then discuss how the resulting ground state sector degeneracies are entirely generated by physical lattice symmetries acting on both the plaquette fluxes and non-contractible Wilson loops. 
Finally, we study the behaviour of the finite-size two-vison gap and matter fermion correlations. We find that the gap does not decrease monotonically with increasing system size, while one still recovers the expected algebraic decay of the correlations at the isotropic point, even for the considered modest system sizes.

\end{abstract}

\maketitle
\section{Introduction}

The study of lattice models on negatively curved geometries has developed rapidly in condensed matter theory over the recent years. 
%
%
Works focusing on single particle physics established how familiar notions of Bloch theory and crystallography must be reformulated on hyperbolic lattices, where the relevant translation groups are generally non-Abelian and their representation theory is correspondingly richer \cite{MaciejkoRayan2021HyperbolicBandTheory,MaciejkoRayan2022AutomorphicBloch,Boettcher2022CrystallographyHyperbolic,Lenggenhager2023NonAbelianHyperbolic}.
This has led to a broad exploration of single particle and topological phenomena in these systems, including non-Abelian semimetals \cite{Tummuru2024HyperbolicNonAbelianSemimetal}, flat bands
\cite{BzdusekMaciejko2022FlatBandsHyperbolic,Mosseri2022ABCagesHyperbolic}, Chern phases and topological gaps \cite{Urwyler2022HyperbolicTopologicalBandInsulators,Liu2022ChernInsulatorHyperbolic,Chen2023HyperbolicHaldaneModels,LuxProdan2023ConvergingPBC}, Anderson localization \cite{Chen2024AndersonHyperbolic}, and topological linear response
\cite{Sun2024TopologicalLinearResponseHyperbolic}, as well as experimental realizations involving circuit QED \cite{Kollar2019HyperbolicCircuitQED}, photonic \cite{Huang2024HyperbolicPhotonicTopologicalInsulators} and topoelectrical systems \cite{Bao_2019_PRB, Lenggenhager2022HyperbolicDrum, Zhang2022TopologicalStatesHyperbolic, Peacock2023HyperbolicMatterCircuits}.
Going beyond the single-particle scenario, another class of problems investigated on hyperbolic lattices is that of spin models, with early studies going back to the 1990s focusing on thermodynamics of Ising models \cite{Series_1990_CommMathPhys, Rietman_1992_JPhysA} (cf. also \cite{Saito_1990}) followed by numerical investigations of phase transitions of Ising and Heisenberg models \cite{Ueda_2007_JPhysSocJapan, Krcmar_2008_JPhysAMathTheor, Baek_2009_PRE, Gendiar_2012_PhysRevE, Gendiar_2014_PhysRevE, Daniska_2016_JPhysA}.

Within this broader development, the Kitaev model on so-called $\{p,q\}$ lattices, which consist of a regular tessellation of the hyperbolic plane by $p$-sided polygons with $q$ edges meeting at each vertex, provides a particularly controlled setting as it is amenable to a quadratic fermionic Hamiltonian in a given gauge background \cite{Kitaev2006Anyons}.
In \cite{Lenggenhager2023NonAbelianHyperbolic,Lenggenhager2025HyperbolicSpinLiquids}, Lenggenhager \textit{et al.} studied the phase diagram for the infinite $\{8,3\}$ lattice using hyperbolic band theory and finite-index normal subgroup sequences and used Lieb's argument to identify the homogeneous $\pi$-flux as the ground state configuration similarly to the Euclidean case \cite{Kitaev2006Anyons}.

Other $\{p,q\}$ lattices can display qualitatively different behavior. For odd $p$, time reversal exchanges conjugate plaquette-flux sectors, allowing the ground state to break time-reversal symmetry spontaneously. Chiral spin-liquid phases have consequently been found on the $\{7,3\}$ and $\{9,3\}$ lattices \cite{Vidal_2025,Dusel_2025,iqbal25}. 

Most of the above studies have been focusing on thermodynamically large system sizes.
Finite systems, however, require dedicated treatment. 

Implementing periodic boundary conditions (PBC) in Euclidean geometries is straightforward as it consists of identifying opposite boundaries in each of the cartesian directions, causing the lattice to wrap around a genus-1 torus. In contrast, implementing PBC in hyperbolic space is non-trivial and results in lattices wrapping around genus-$g$ surfaces instead.
In addition to spin models, appropriate handling of boundary conditions is necessary when dealing with hyperbolic quantum error correcting codes \cite{TillichZemor2009QuantumLDPC,BreuckmannTerhal2016HyperbolicSurfaceCodes,BreuckmannEberhardt2021QuantumLDPC,HiggottBreuckmann2024HyperbolicFloquetCodes,Fahimniya2025HyperbolicFloquetCodes,Boettcher2020QuantumSimulationHyperbolic,Pastawski2015HaPPY}. Beyond these, constructions of hyperbolic tensor networks with the scope of investigating holography have also been developed \cite{Jahn2022TensorNetworkAdSqCFT,Steinberg2023HyperinvariantCodes}.

In this context, a detailed study of the Kitaev model of moderate finite size on hyperbolic lattices with PBC is the main focus of this paper. This is motivated by possible experimental realizations using current quantum hardware, such as superconducting circuits, neutral atoms or trapped ions, which often operate dozens or hundreds of spin carriers (qubits) \cite{Vezvaee2026SurfaceCodeScaling, Semeghini2021TopologicalSpinLiquidsSimulator, Ransford2026Helios98Qubits} and where spin models can be realized either in a digital or analog fashion \cite{Blais_2021_RMP, Pritchard2026NeutralAtomQuantumComputing, FossFeig2025ProgressTrappedIonQuantumSimulation, Monroe_2021_RMP, Katz2025HybridDigitalAnalog}. For instance, arbitrary connectivities between the spins, here realizing the hyperbolic lattice, can be achieved in trapped ion systems \cite{Korenblit2012ArbitraryLatticeIons} and efficient protocols for generation of fermionic Gaussian states, i.e. the ground states of the matter fermions of the Kitaev model, have been developed \cite{Jiang_2018_PRApp, Verstraete_2009_PRA}.

While lattice models on hyperbolic spaces have been by now studied in some detail, we find it helpful to discuss the necessary notions in a self-contained manner. For this reason, in Sec. \ref{sec:GroupTheory}, \ref{sec:HyperbolicGeometry} we provide a review of triangle groups and their presentations and hyperbolic geometry together with a number of illustrative examples throughout the manuscript. Exploiting the group theoretic construction in Sec. \ref{sec:LatticeConstructions} we provide a pseudocode which generates the adjacency matrix of an arbitrary $\{p,q\}$ lattice with consistent PBC, employing the concept of regular maps.
In doing so, we do not resort to explicit construction of non-Abelian translation groups \cite{MaciejkoRayan2021HyperbolicBandTheory,MaciejkoRayan2022AutomorphicBloch,Boettcher2022CrystallographyHyperbolic,Lenggenhager2023NonAbelianHyperbolic}. Our construction starts directly from an arbitrary finite quotient and reconstructs its complete compact cellulation, incidence structure, homology and Wilson sectors without first choosing a Bravais cell or a coherent supercell sequence.
Sec. \ref{sec:homology} then discusses how to extract the corresponding plaquette structure, homology cycles and Wilson sectors. Equipped with the necessary ingredients we define the Kitaev model in Sec. \ref{sec:kitaev_model}, \ref{sec:parity_projection}. Finally, in Sec. \ref{sec:Results} we perform an extensive analysis of the model on $\{8,3\}$ and $\{12,3\}$ geometries and moderate system sizes up to 72 lattice sites. Specifically, for each studied quotient, we enumerate all gauge invariant superselection sectors, solve the quadratic Majorana Hamiltonian and impose the physical parity projection independently in every sector. We therefore make no prior assumption about the ground state flux background. This allows us to determine the exact physical ground state of each finite PBC cluster, quantify its deviation from the homogeneous $\pi$ flux configuration, resolve the role of parity projection, identify the symmetry origin of its ground state sector degeneracy, and define vison excitations relative to the complete ground state flux manifold. In this way, the finite quotient construction provides a controlled setting in which geometry, topology and the physical constraints of the Kitaev representation can be treated on equal footing. We discuss open questions and future directions in Sec. \ref{sec:Conclusion}.
~\\




\section{Group theory primer}
\label{sec:GroupTheory}
Here we recall the group-theoretical procedure how to construct a lattice with periodic boundary conditions (PBC), regardless of the underlying space, by introducing group presentations, normal subgroups, cosets and quotient groups \cite{Rotman1995TheoryOfGroups}.\\ 

We start by introducing a compact way of stating a group's structure. 
\begin{definition}
A \textit{group presentation} specifies a group in terms of its generators $a_i$ and defining relators $r_i$, namely $G=\langle a_1,\dots,a_l \mid r_1,\dots,r_m\rangle$, where each relator specifies a constraint $r_i=\mathds{1}$, with $\mathds{1}$ the identity element. 
\end{definition}
\begin{example}
A simple example is given by the cyclic group of order $n$, with presentation $\langle a\mid a^n\rangle$. This notation implies that applying the generator $a$ for $n$ times yields the identity, $a^n=\mathds{1}$, making $n$ the \textit{order} of the group.
\end{example}
\begin{definition}
A subgroup $N$ of a group G is called \textit{normal}, denoted $N\triangleleft G$, if it is invariant under conjugation by elements of $G$, i.e. $gNg^{-1}=N$ for all $g\in G$. 
\end{definition}
\begin{definition}
A \textit{left coset} of a subgroup $N$ is the set $gN=\{gn:n\in N\}$, while a \textit{right coset} is $Ng=\{ng:n\in N\}$. 
\end{definition}
\begin{definition}
    The index of a subgroup $N$ in a group $G$ is the number of left (right) cosets of $N$ in $G$. It is denoted as $[G:N]$.
\end{definition}
\begin{definition}
    A group is \textit{torsion-free} if its only finite-order element is the identity. 
\end{definition}
\begin{remark}
When $N\triangleleft G$ is normal, left and right cosets coincide, $gN=Ng$, and these cosets form the elements of the \textit{quotient group} $G/N$; we denote them by $[g]$.
\end{remark} 
We will now elucidate how this is relevant in the context of defining PBC. In the Euclidean case, one typically starts from a free abelian group $G$ representing a discrete Euclidean symmetry, most often translations. One then chooses a normal subgroup $N\triangleleft G$ and forms the quotient group $Q=G/N$.
Elements of $Q$ are equivalence classes of elements of $G$, where two elements are identified if they differ by an element of $N$. This quotienting operation is precisely what implements periodic boundary conditions.

\begin{example} 
The simplest example of the quotienting procedure is a one-dimensional lattice with PBC. The infinite chain is generated by the translation group $G=\mathbb Z=\{\ldots,-2,-1,0,1,2,\ldots\}$, where $n\in\mathbb Z$ denotes translation by $n$ lattice spacings. To obtain a periodic chain with $L=4$ sites, for example, we impose the equivalence relation $n\sim n+4$ by taking a quotient by the subgroup $N=4\mathbb Z=\{\ldots,-8,-4,0,4,8,\ldots\}$. Thus $Q=\mathbb Z/4\mathbb Z$. The elements of this quotient are the four cosets $[0]=4\mathbb Z$, $[1]=1+4\mathbb Z$, $[2]=2+4\mathbb Z$, and $[3]=3+4\mathbb Z$. Addition of translations is inherited modulo four, $[a]+[b]=[a+b]\mod 4$, and, consequently, $Q \simeq \mathbb Z_4$.
\end{example}
An analogous procedure applies in two Euclidean dimensions, where the translation group is $G=\mathbb Z^2$. Quotienting by a two-dimensional sublattice produces a finite periodic cluster embedded on a torus. The peculiarity of the hyperbolic construction in two dimensions is that the starting group is no longer the abelian translation group $\mathbb Z^2$, but a triangle group or one of its orientation-preserving subgroups. After quotienting by a suitable normal subgroup, the resulting compact surface need not have genus $g=1$. We now move on to describing the basics of hyperbolic geometry and how to relate the abstract concept of a group to a lattice with PBC.

\section{Hyperbolic geometry primer}
\label{sec:HyperbolicGeometry}

We start from the hyperbolic space of dimension $2$, $\mathbb{H}^2$ and consider its Poincaré disk representation $\mathbb D$ as the set of complex numbers inside the unit disk,
\begin{equation}
\mathbb{D}=\{z\in\mathbb{C}\,:\,|z|<1\}.
\eeq

In this model, distances are measured with the hyperbolic metric
\beq
ds^2=\frac{4|dz|^2}{(1-|z|^2)^2},
\eeq
see \cite{BalazsVoros1986ChaosPseudosphere,SaussetTarjus2007PBCPseudosphere} for details. This expression shows that distances increase rapidly as one approaches the boundary: although the disk has finite radius in Euclidean terms, the boundary $|z|=1$ lies at infinite distance with respect to this metric. As a result, the entire infinite hyperbolic plane is represented inside a finite disk. The symmetries of the hyperbolic plane are transformations that preserve this metric. The orientation-preserving ones form the group $\mathrm{PSU}(1,1)$, which acts on the disk through Möbius transformations of the form
\begin{equation}
z\mapsto \frac{az+b}{b^*z+a^*},\qquad a,b\in\mathbb{C},\qquad |a|^2-|b|^2=1.
\end{equation}

Certain discrete subgroups of these isometries generate regular tilings of the hyperbolic plane. 
\begin{definition} 
A \textit{tiling} of a surface is a decomposition of the surface into non-overlapping polygons that cover it completely. 
\end{definition}
\begin{definition}
A tiling is called \textit{regular} if all polygons are congruent and the same number of faces meet at every vertex, so that the local structure is identical throughout the constructed surface.
\end{definition}
These regular tilings provide the infinite geometric backbone from which our lattices will be constructed. At this stage, however, the tiling still extends over the entire hyperbolic plane. The construction of finite periodic lattices will require us to identify suitable regions of this infinite tiling, as discussed below.
A systematic way to construct regular tilings of the hyperbolic plane is provided by \textit{triangle groups}. 
\begin{definition}
Consider a geodesic triangle with interior angles $
(\frac{\pi}{r},\,\frac{\pi}{q},\,\frac{\pi}{p})$, $r,q,p \in {\mathbb N}$, as the one in Fig.\ref{fig_fundamental_triangle}(c).
\begin{figure}[h]
    \centering
    \includegraphics[width=1\linewidth]{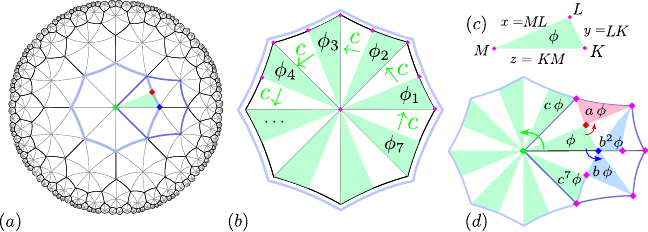}
    \caption{(a) Example of the $\{8,3\}$ tessellation generated by reflections through the edges of a hyperbolic fundamental domain (green geodesic triangle) with internal angles $\pi/2$, $\pi/3$ $\pi/8$ incident on red, blue and green vertices respectively. (b) Faces are identified as orbits of the generator $c=zx$. (c) A pictorial representation of a flag, an oriented triple specified by a vertex, an edge and a face, here bounded by points $K,L,M$. A $(2,3,8)$ tessellation is generated by reflections along the sides $x,y,z$ of this triangle with internal angles $(\pi/2,\pi/3,\pi/8)$. (d) A zoom-in on the region surrounded by blue lines in panel (a), showing the explicit rotational action of the orientation-preserving generators $a,b,c$ on a flag $\phi$.} 
    \label{fig_fundamental_triangle}
\end{figure}
The \textit{triangle group} $\Delta(r,q,p)$ is the group generated by reflections in the three sides of this triangle; for this reason it is a Coxeter group. Concretely, it admits the presentation:
\begin{align}
\Delta&(r,q,p)=\notag\\
=&\langle x,y,z \mid x^2=y^2=z^2=(xy)^r=(yz)^q=(zx)^p=1\rangle .
\end{align}
Here $x,y,z$ denote reflections across the three sides of the triangle, $x^2=y^2=z^2=1$ expresses that reflections are involutions, and $(xy)^r=1$ means that the product of two reflections across adjacent sides is a rotation of order $r$ (and similarly for $(yz)^q$ and $(zx)^p$). 
\end{definition}
The subgroup of orientation-preserving elements is called the \textit{proper triangle group} (or Von Dyck Group) and is denoted $\Delta^+(r,q,p)$. Since each reflection reverses orientation, while a product of two reflections preserves it, $\Delta^+(r,q,p)$ consists precisely of all even-length applications of the generators $x,y,z$. It is therefore an \textit{index-two} subgroup, meaning that it contains exactly half of the elements of $\Delta(r,q,p)$. To express this, we use the index notation $ [\Delta(r,q,p):\Delta^+(r,q,p)] = 2$.
Moreover, it is a normal subgroup, $\Delta^+(r,q,p)\;\triangleleft\;\Delta(r,q,p)$.
When the condition
\beq
\frac{1}{r}+\frac{1}{q}+\frac{1}{p}<1
\eeq
is satisfied, the triangle tiles the hyperbolic plane. The action of the triangle group generators creates a regular tiling by congruent copies of this triangle.
\begin{figure}[h!]
    \centering
    \includegraphics[width=\linewidth]{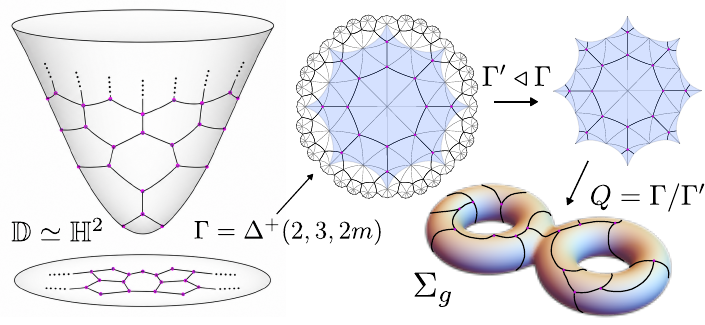}
    \caption{Schematic construction of a finite periodic hyperbolic cluster. 
Starting from the infinite hyperbolic plane $\mathbb H^2$, represented by the Poincaré disk $\mathbb D\simeq \mathbb H^2$, we consider the infinite regular tiling generated by the orientation-preserving triangle group $\Gamma=\Delta^+(2,3,2m)$.  Periodic boundary conditions are imposed by choosing a finite-index normal subgroup $\Gamma'\triangleleft\Gamma$. The shaded region in the disk represents a fundamental domain for this subgroup. Identifying its boundary according to the action of $\Gamma'$ produces a compact genus-$g$ surface $\Sigma_g=\mathbb H^2/\Gamma'$. The finite quotient group  $Q=\Gamma/\Gamma'$ labels the cosets (oriented flags) from which the vertices, edges and faces of the PBC cluster are reconstructed.}
\label{fig_compactification_procedure}
\end{figure}
In this work we employ the orientation-preserving subgroup. 
\begin{definition}The \textit{orientation-preserving triangle group} is the one with presentation
\begin{equation}
\Delta^+(r,q,p)=\langle a,b,c \mid a^r=b^q=c^p=abc=1\rangle.
\label{eq:orientation_preserving_triangle_group}
\end{equation}
\end{definition}
Here $a,b,c$ are orientation-preserving rotations about the three vertices of the fundamental triangle generated by composition of the reflections $a=xy, \;b=yz, \;c=zx$ (see fig. \ref{fig_fundamental_triangle} (d)). Their orders are fixed by the triangle's angles: $a$ has order $r$, $b$ has order $q$, and $c$ has order $p$. The relation $abc=1$ states that the product of the three local rotations around the triangle is trivial. 

In this work we consider $(r,q,p) = (2,3,2m)$, $m \in {\mathbb N}$; $q=3$ and $p$ even are required for trivalence and bipartiteness, respectively. These two properties are needed for defining the models we study.
For $r=2$, $\Delta^+(2,q,p)$ corresponds to a regular tiling of Schläfli type $\{p,q\}$. Thus $q=3,p=2m$ gives $\{2m,3\}$. We will use these notations interchangeably. For this family this becomes
\begin{equation}
\Delta^+(2,3,2m)= \langle a,b,c \mid a^2=b^3=c^{2m}=abc=1\rangle,
\end{equation}
again with $a=xy, \;b=yz, \;c=zx$. This presentation makes explicit that $\Delta^+(r,q,p)$ is generated purely by rotations and acts by orientation-preserving isometries of the hyperbolic plane. In particular, $\Gamma\equiv\Delta^+(2,3,2m)$ determines the local geometry and symmetry of the infinite $\{2m,3\}$ tiling underlying our lattices.

Our numerical calculations, however, require finite lattices with PBC. To obtain them, we quotient this infinite geometric structure by a suitable finite-index subgroup $\Gamma'\triangleleft \Gamma$.  
In the next section we show explicitly how the quotient of the triangle group determines the finite PBC lattice.

\section{Lattice construction through regular maps}
\label{sec:LatticeConstructions}

We now make precise the passage introduced above from the infinite hyperbolic tiling to the finite PBC lattice. Similar constructions appear in the mathematical and QEC literature; see, for example, \cite{BreuckmannTerhal2016HyperbolicSurfaceCodes,rayanerrorcorrection}. Let $\Gamma=\Delta^+(2,3,2m)$ be the orientation-preserving triangle group generating the infinite $\{2m,3\}$ tiling. Choosing an appropriate normal subgroup $\Gamma'\triangleleft\Gamma$ identifies points of the tiling and makes it descend to a finite decomposition of the compact, orientable surface $\Sigma_g=\mathbb H^2/\Gamma'$.  Algebraically, the corresponding finite quotient group is
\begin{equation}
Q=\Gamma/\Gamma'.
\label{eq:quotient}
\end{equation} 
Following Conder \cite{ConderRegularMapsLists}, we denote such normal subgroups $\Gamma'$ as $\mathrm{Tg.n}$, where $g$ denotes the genus of the resulting compact surface and $n$ distinguishes inequivalent subgroups at fixed genus. The subgroups are specified by adding one or more additional relators, or \textit{words}, for which we use the notation $\bar w=(w_1,\dots,w_k)$, to the presentation of $\Gamma$. A number of low-index normal subgroups of triangle groups were enumerated by Conder using the LINS algorithm \cite{LINS0.9}. We define
\begin{equation}
\Gamma'=\langle\!\langle \bar w\rangle\!\rangle\triangleleft\Gamma,
\end{equation}
where $\langle\!\langle \bar w\rangle\!\rangle$ denotes the normal closure of $\bar w$, namely the smallest normal subgroup containing $\bar w$, and restrict to choices for which $\Gamma'$ is finite-index and torsion-free. The torsion-free condition is required so that the result of the quotienting procedure is a smooth surface. Different choices of relators $\bar w$ can produce inequivalent finite quotients, corresponding to different compact surfaces and different finite tilings. In particular, the size of $Q$, given by $|Q|=[\Gamma:\Gamma']$, and the global topology, encoded by the genus of the surface, can change. 

\begin{example}
The quotient T2.1 is obtained from $\Delta^+(2,3,8)$ and is defined by $\Gamma_{\mathrm{T2.1}}=\langle\!\langle \bar w\rangle\!\rangle$ with $\bar{w} = \{ w \}$, $w=c\,b\,a\,c\,b^{-1}\,c^{-1}\,a\,c$ and $Q=\Delta^+(2,3,8)/\Gamma'_{\mathrm{T2.1}}$. It has $|Q|=48$ oriented flags (cf. Definition.~\ref{def:OrientedFlag}). 
\end{example}

\begin{example}
The quotient T4.1 is obtained from $\Delta^+(2,3,12)$ and is defined by $w_1=a\,c\,b\,a\,c\,b\,a\,b^{-1}\,c^{-1}$, $w_2=a\,c^3\,b^2\,c^{-4}$, so $\bar w = (w_1, w_2)$, $\Gamma'_{\mathrm{T4.1}}=\langle\!\langle \bar w\rangle\!\rangle$, $Q=\Delta^+(2,3,12)/\Gamma'_{\mathrm{T4.1}}$, $|Q|=72$. 
\end{example}
It is now useful to introduce some objects that allow us to rigorously define maps (cf. Chapter 0 of \cite{Hatcher2002AlgebraicTopology}).
\begin{definition}
A \textit{homeomorphism} is a continuous bijection whose inverse is also continuous.
\end{definition}
\begin{definition}
Let
\begin{equation*}
D^n=\{x\in\mathbb{R}^n:\|x\|\leq 1\}.
\end{equation*}
A \textit{topological $n$-disk} is a topological space homeomorphic to $D^n$.
\end{definition}

\begin{definition}
An $n$-\textit{cell} is a topological space homeomorphic to the open disk
\begin{equation*}
{D}_{\circ}^n=\{x\in\mathbb{R}^n:\|x\|<1\}.
\end{equation*}
In particular:
\begin{enumerate}
    \item a $0$-cell is a point;
    \item a $1$-cell is homeomorphic to an open interval (edge);
    \item a $2$-cell is homeomorphic to an open disk (face).
\end{enumerate}
\end{definition}
As anticipated, after quotienting $\Gamma/\Gamma'$, the infinite tiling descends to a map \cite{MoharThomassen2001GraphsOnSurfaces,GrossTucker1987TopologicalGraphTheory,EdmondsEwingKulkarni1982TorsionFree,ConderRegularMapsLists,ConderDobcsanyi2001RegularMapsSmallGenus,Grunbaum1994HollowFaces,conderRegularMapsNonorientable1995} $\mathcal M_Q$ on the compact surface $\Sigma_g=\mathbb H^2/\Gamma'$. 
\begin{definition}
A \textit{map} $\cal M$ is an embedding of a graph onto a closed surface without boundary. The $2$-cells of the map are the graph's faces, $1$-cells are edges, and $0$-cells are vertices.
\end{definition}
Maps are of type $\{p,q\}$ when every face has $p$ edges and every vertex has valence $q$; the maps considered here are therefore of type $\{2m,3\}$. 
\begin{definition}
\label{def:OrientedFlag}
An \textit{oriented flag} is a mutually incident vertex-edge-face triple compatible with a choice of local orientation.
\end{definition}

This means that the flag is comprised of a vertex $v$ which is an endpoint of an edge $e$, and $e$ lies on the boundary of a face $f$. We denote it by $\phi$ and in Fig.~\ref{fig_fundamental_triangle}(c) we represent it as the region bounded by points $(K,L,M)$. The flag is oriented because one can associate with the green triangles a consistent orientation. When the triangle is then reflected along any of the $x,y,z$ edges the orientation of the flag is reversed (white triangles in Fig.~\ref{fig_fundamental_triangle}(b)).

\begin{definition}
An \textit{automorphism} of a map $\mathcal M$ is a permutation of its flags that preserves all incidence relations amongst them. The group of all such automorphisms is denoted by $\mathrm{Aut}\,\mathcal M$.
\end{definition}

\begin{example}
Consider the map obtained from the vertices, edges, and faces of a cube. A rotation by $90^\circ$ about an axis passing through the centers of two opposite faces permutes its vertices, edges, and faces while preserving the incidence relations between them. It therefore induces a permutation of its cells and defines an automorphism of the map. Thus automorphisms can be understood as the combinatorial symmetries of the lattice.
\end{example}
\begin{definition}
A map is called \textit{regular} if $\mathrm{Aut}\,\mathcal M$ acts transitively on its flags.
\end{definition}
\begin{definition}
A map is \textit{orientable} if the surface on which it is embedded is orientable. 
\end{definition}
\begin{definition}
A map is \textit{reflexible} if it admits an orientation-reversing automorphism (a reflection).
\end{definition}
We restrict to orientable, reflexible regular maps. The latter condition is needed because we don't consider chiral maps. For maps with the listed properties, the orientation-preserving automorphisms form a normal subgroup of index two,
\begin{equation}
\mathrm{Aut}^+\,\mathcal M\triangleleft\mathrm{Aut}\,\mathcal M,
\qquad
[\mathrm{Aut}\,\mathcal M:\mathrm{Aut}^+\,\mathcal M]=2.
\end{equation}
$\mathrm{Aut}^+\,\mathcal M$ is also called the \textit{orientation-preserving} or \textit{rotation subgroup}. It acts transitively on the oriented flags.
$Q=\Gamma/\Gamma'$ is the finite quotient group whose elements label the oriented flags of $\mathcal M_Q$. In the present construction, the orientation-preserving automorphism group is realized concretely by
\begin{equation}
Q\simeq \mathrm{Aut}^+\,\mathcal M_Q,
\end{equation}
where $\simeq$ denotes an isomorphism.
Since there is a one-to-one correspondence between the oriented flags and the elements of $Q=\Gamma/\Gamma'$, which are cosets $[g]$ of $\Gamma'$ in $\Gamma$, we label each flag by their corresponding group element to create the flag set
\begin{equation}
\Phi_Q=\{\phi_{[g]}:[g]\in Q\},
\qquad [g]=\Gamma'g=g\Gamma'.
\end{equation}
Following the right-coset convention of Conder \cite{ConderDobcsanyi2001RegularMapsSmallGenus,conderRegularMapsNonorientable1995}, the group
$Q\simeq\operatorname{Aut}^{+}M_Q$ acts on $\Phi_Q$ by right multiplication:
\begin{equation}
    \phi_{[g]}\cdot[h]:=\phi_{[gh]},
    \qquad [g],[h]\in Q.
\end{equation}
To construct the cells, left multiplication by the images of $a$, $b$, and $c$ defines permutations $\pi_a,\pi_b,\pi_c\in S_{|Q|}$:
\begin{equation*}
\pi_a(\phi_{[g]})=\phi_{a[g]},\qquad
\pi_b(\phi_{[g]})=\phi_{b[g]},\qquad
\pi_c(\phi_{[g]})=\phi_{c[g]}.
\end{equation*}
These permutations commute with the right action of $Q$.

These correspond respectively to an order-$2$ rotation about an edge midpoint, an order-$3$ rotation about a vertex, and an order-$2m$ rotation about a face.

For $j\in\{a,b,c\}$, let $\langle j\rangle$ denote the cyclic subgroup generated by $j$,
so that
\begin{align}
\langle {a}\rangle&=\{1,a\},\\
\langle {b}\rangle&=\{1,b,b^2\},\\
\langle {c}\rangle&=\{1,c,c^2,\ldots,c^{2m-1}\}.
\end{align}
The corresponding orbit is
\begin{equation}
\mathrm{Orb}_{\langle\pi_j\rangle}(\phi_{[g]})= \{\phi_{h[g]}:h\in\langle j\rangle\}.
\end{equation}
So under the identification $\Phi_Q\simeq Q$, each orbit is identified with the
right coset
\begin{equation}
\langle j\rangle[g]=
\{h[g]:h\in\langle j\rangle\}.
\end{equation}
The sets of edges, vertices, and faces of the map $\cal M_Q$ in terms of cosets are therefore
\begin{align}
\textcolor{red}{\mathcal E_Q}
&=\{\langle{a}\rangle[g]:[g]\in Q\},
\\
\textcolor{blue}{\mathcal V_Q}
&=\{\langle{b}\rangle[g]:[g]\in Q\},
\\
\textcolor{green}{\mathcal F_Q}
&=\{\langle{c}\rangle[g]:[g]\in Q\}.
\end{align}
We identify these cosets with the corresponding $n$-cells of $\mathcal M_Q$. Incidence is defined by non-empty intersection of the corresponding cosets in $Q$. For example, a vertex coset $\langle b\rangle[g]$ and an edge coset $\langle a\rangle[g]$ are incident if and only if
\begin{equation}
\langle b\rangle[g]
\cap
\langle a\rangle[g]
\neq\varnothing,
\end{equation}
and analogously for vertex-face and edge-face incidence. 

\begin{example}
In Fig.\ref{fig_fundamental_triangle}(b), we identify a face as the orbit of a flag $\phi$ under the generator $c$ as $F=\mathrm{Orb}_{\langle\pi_c\rangle}(\phi)=\{\phi_1, \,c\,\phi_1=\phi_2,c^2\,\phi_1=\phi_3,\dots,c^7\,\phi_1=\phi_8\}$.
\end{example}
\begin{example}
In Fig.\ref{fig_fundamental_triangle}(d), the action of \textcolor{red}{a} identifies edges incident on $\phi$ ($E_\phi=\mathrm{Orb}_{\langle\pi_a\rangle}(\phi)=\{\phi,a\,\phi \}$, order 2), the action of \textcolor{blue}{b} identifies vertices incident on $\phi$ ($V_\phi=\mathrm{Orb}_{\langle\pi_b\rangle}(\phi)=\{\phi,b\,\phi,b^2\,\phi \}$, order 3), the action of \textcolor{green}{c} identifies faces incident on $\phi$ ($F=\mathrm{Orb}_{\langle\pi_c\rangle}(\phi)=\{\phi, \,c\,\phi=\phi_1,c^2\,\phi=\phi_2,\dots,c^7\,\phi\}$, order 8). Grouping all orbits generates the set of edges, vertices and faces of the resulting lattices.
\end{example}

We call the quotient \textit{non-degenerate} if the images of $a$, $b$, and $c$ in $Q$ have orders $2$, $3$, and $2m$, respectively. For a non-degenerate quotient, the orders of the three generators are thus preserved, therefore
\begin{equation}
|\textcolor{red}{\mathcal E_Q}|=\frac{|Q|}{2},\qquad
|\textcolor{blue}{\mathcal V_Q}|=\frac{|Q|}{3},\qquad
|\textcolor{green}{\mathcal F_Q}|=\frac{|Q|}{2m}.
\end{equation}
\begin{definition}
Given a graph $\textbf{G}$, the vertex, edge and face sets $V(\textbf{G}),E(\textbf{G}),F(\textbf{G})$ are defined as the collections of vertices, edges and faces generated by the embedded graph itself.
\end{definition}
For the graph $\textbf{G}$ generated by $\mathcal M_Q$ we have $|\mathcal V_Q|=|V(\textbf{G})|=V$ vertices, $|\mathcal E_Q|=|E(\textbf{G})|=E$ edges, and $|\mathcal F_Q|=|F(\textbf{G})|=F$ faces. These will be identified with the sites, bonds, and plaquettes of the PBC cluster. The genus of the surface in which $\textbf{G}$ is embedded is obtained from the usual Euler characteristic formula:
\begin{equation}
\chi=V-E+F=2-2g.
\label{eq:graph_genus}
\end{equation}
Thus, the full lattice structure is encoded by the action of $Q$ on oriented flags, with the generators $\textcolor{red}{a}$, $\textcolor{blue}{b}$, and $\textcolor{green}{c}$ generating the edge, vertex, and face orbits, respectively.
This gives a purely combinatorial PBC construction from which the Kitaev model connectivity, plaquette fluxes, homology generators and Wilson sectors can be extracted algorithmically. 

\begin{example}
Consider the quotient T2.1, which gives a regular map of type $\{8,3\}$. In this case $|Q|=48$, while the generators $a$, $b$, and $c$ have orders $2$, $3$, and $8$, respectively. Hence
\begin{equation}
E=\frac{48}{2}=24,\qquad
V=\frac{48}{3}=16,\qquad
F=\frac{48}{8}=6.
\end{equation}

Consider first the cosets containing the identity element $[1]\in Q$, then
\begin{align}
e_0&=\langle a\rangle[1]=\{[1],[a]\}\in\textcolor{red}{\mathcal E_Q},\\
v_0&=\langle b\rangle[1]=\{[1],[b],[b^2]\}\in\textcolor{blue}{\mathcal V_Q},\\
f_0&=\langle c\rangle[1]=\{[1],[c],[c^2],\ldots,[c^7]\}\in\textcolor{green}{\mathcal F_Q}.
\end{align}
Since $[1]\in e_0$, $[1]\in v_0$, $[1]\in f_0$, we have
\begin{equation}
e_0\cap v_0\neq\varnothing,\qquad
e_0\cap f_0\neq\varnothing,\qquad
v_0\cap f_0\neq\varnothing,
\end{equation}
so these three cells are mutually incident. The other element of the edge coset satisfies $[a]\in e_0$ and it also belongs to
\begin{align}
v_1&=\langle b\rangle[a]
=\{[a],[ba],[b^2a]\},\\
f_1&=\langle c\rangle[a]
=\{[a],[ca],[c^2a],\ldots,[c^7a]\}.
\end{align}
Hence
\begin{equation}
[a]\in e_0,\qquad [a]\in v_1,\qquad [a]\in f_1,
\end{equation}
and therefore
\begin{equation}
e_0\cap v_1\neq\varnothing,\qquad
e_0\cap f_1\neq\varnothing,\qquad
v_1\cap f_1\neq\varnothing.
\end{equation}
Thus $e_0$ is incident to both $v_0$ and $v_1$, and to both $f_0$ and $f_1$. The two elements $[1]$ and $[a]$ of the edge coset therefore identify its two endpoints and its two incident faces. The same construction makes the local coordination explicit. Since
\begin{equation}
[1],[b],[b^2]\in v_0,
\end{equation}
these elements belong respectively to the edge cosets
\begin{equation}
\langle a\rangle[1],\qquad
\langle a\rangle[b],\qquad
\langle a\rangle[b^2].
\end{equation}
This implies that
\begin{equation}
v_0\cap\langle a\rangle[1]\neq\varnothing,\qquad
v_0\cap\langle a\rangle[b]\neq\varnothing,\qquad
v_0\cap\langle a\rangle[b^2]\neq\varnothing.
\end{equation}
These are therefore the three edges incident to $v_0$, showing that every vertex is trivalent.

Likewise,
\begin{equation}
[1],[c],[c^2],\ldots,[c^7]\in f_0,
\end{equation}
and these elements belong respectively to
\begin{equation}
\langle a\rangle[1],\,
\langle a\rangle[c],\,
\langle a\rangle[c^2],\,
\ldots,\,
\langle a\rangle[c^7].
\end{equation}
Hence
\begin{equation}
f_0\cap\langle a\rangle[c^k]\neq\varnothing,
\qquad k=0,\ldots,7,
\end{equation}
so these are the eight edges incident to $f_0$, reproducing one of the octagonal faces of the $\{8,3\}$ map.
\end{example}

\begin{example}
For T2.1, since $|Q|=48$, we have $V=\frac{|Q|}{3}=16$,
$E=\frac{|Q|}{2}=24$ and $F=\frac{|Q|}{8}=6$. It follows from Eq.~(\ref{eq:graph_genus}) that $\chi=V-E+F=-2$ so $g=2$.
\end{example}

\begin{example}
For T4.1, since $|Q|=72$, we have $V=\frac{|Q|}{3}=24$,
$E=\frac{|Q|}{2}=36$ and $F=\frac{|Q|}{12}=6$, $\chi=V-E+F=-6$ so $g=4$.
\end{example}

\begin{remark}
$\mathrm T2.1$ is isomorphic to Grünbaum's orientable genus-2 regular map $\mathrm W\#24.21$ \cite{Grunbaum2003AreYourPolyhedra}.
\end{remark}
For each quotient, we check explicitly that the order of the permutations is
\begin{equation}
\mathrm{ord}({\color{red}\pi_a})=2,\qquad
\mathrm{ord}({\color{blue}\pi_b})=3,\qquad
\mathrm{ord}({\color{green}\pi_c})=2m,
\end{equation}
ensuring that the quotient is non-degenerate and preserves the local $\{2m,3\}$ structure. 
The above construction determines the incidence relations and thus the nearest-neighbour connectivity of the underlying graph $\mathbf G$. 
\begin{algorithm}[H]
\caption{Construction of the adjacency matrix from the quotient cosets}
\label{alg:adjacency-matrix}
\begin{algorithmic}
\Statex \textbf{Input:} Distinct vertex cosets
$\mathcal V_Q=\{v_i=\langle b\rangle[g_i]\}_{i=1}^{V}$,
where $V=|Q|/3$, together with the edge cosets $\mathcal E_Q$
\Statex \textbf{Output:} Adjacency matrix $\mathrm{Adj}(\mathbf G)$
\State $\mathrm{Adj}(\mathbf G)\gets 0_{V\times V}$
\ForAll{$e\in\mathcal E_Q$}
    \State $I_e\gets
    \{i\in\{1,\ldots,V\}:v_i\cap e\neq\varnothing\}$
    \ForAll{$i,j\in I_e$ with $i<j$}
        \State $\mathrm{Adj}_{ij}(\mathbf G)\gets 1$
        \State $\mathrm{Adj}_{ji}(\mathbf G)\gets 1$
    \EndFor
\EndFor
\State \textbf{return} $\mathrm{Adj}(\mathbf G)$
\end{algorithmic}
\end{algorithm}
To summarize, we start from the orientation-preserving triangle group $\Gamma=\Delta^+(2,3,2m)$, we choose a torsion-free finite-index normal subgroup $\Gamma'\triangleleft\Gamma$. The quotient $\Sigma_g=\mathbb H^2/\Gamma'$ is a compact orientable hyperbolic surface, while the finite group $Q$ acts regularly on the oriented flags of the map $\mathcal M_Q$. The action of the three different generators allows to define the sets $\textcolor{red}{\mathcal E_Q},\textcolor{blue}{\mathcal V_Q},\textcolor{green}{\mathcal F_Q}$ which in turn identify the vertex, edge and face sets $V(\textbf{G}),E(\textbf{G}),F(\textbf{G})$ of the graph $\mathbf{G}$ embedded in the orientable surface. We have implemented this procedure in GAP \cite{GAP2025}.\\ 
Once we have defined and connected our PBC cluster, we can move on to understanding the topology of the lattice through homology, describing the three-colored Kitaev model and its Majorana representation, and finally constructing the projection operator to select physical states.

\section{Homology}
\label{sec:homology}

Having constructed a finite lattice represented by a graph $\mathbf{G}$ from the quotient group $Q$, we now characterize its topology. This is essential for classifying gauge sectors that naturally arise in the Kitaev model (see Sec.~\ref{sec:kitaev_model}). The appropriate mathematical framework is provided by discrete homology theory \cite{Hatcher2002AlgebraicTopology,Giblin_2010}. We consider the cellular decomposition $\mathcal M_Q$ yielding a graph $\textbf{G}$ embedded in the surface $\Sigma_g$, and work over $\mathbb Z_2$, anticipating the $\mathbb{Z}_2$ gauge degrees of freedom of the Kitaev model.
\begin{definition}
A 1-chain is a collection of edges specified by a binary vector $x=(x_{1},\dots,x_{E}) \in \mathbb{Z}_2^E$, where an entry $x_e = 1$ indicates that edge $e\in E(\textbf{G})$ is included, and $x_e = 0$ otherwise. 
\end{definition}
\begin{example}
The group $\mathbb Z_2=\{0,1\}$ has addition modulo $2$, so $1+1=0$. Its $n$-fold product
\[
\mathbb Z_2^n=\underbrace{\mathbb Z_2\times\cdots\times\mathbb Z_2}_{n\text{ times}}
\]
consists of binary vectors $(x_1,\ldots,x_n)$ with $x_i\in\{0,1\}$, added componentwise modulo $2$. For example,
\[
\mathbb Z_2^2=\{(0,0),(1,0),(0,1),(1,1)\},
\]
with $(1,0)+(0,1)=(1,1)$ and $(1,1)+(1,1)=(0,0)$. Thus $|\mathbb Z_2^n|=2^n$.
\end{example}
\begin{remark}
The above example illustrates that in $\mathbb{Z}_2$, addition of 1-chains is defined modulo two, so that edges appearing in both sets cancel out while edges appearing in only one are retained. See Fig.\ref{fig_cycle_plaquette} for a visual example.
\end{remark}
To formalize the notion of a closed path on $\textbf{G}$, we define some useful objects.
\begin{definition}
The vertex-edge incidence matrix
\begin{equation}
B \in \mathbb{Z}_2^{V \times E}, 
\;\;\;
B_{v\,e} =
\begin{cases}
1 & \text{if $e\in E(\textbf{G})$ inc. on $v\in V(\textbf{G})$},\\
0 & \text{otherwise}.
\end{cases}
\end{equation}
\end{definition}
\begin{definition}
A 1-chain $x$ is a cycle (it is closed) if 
\begin{equation}
B\,x = 0.
\end{equation}
\end{definition}
The space of closed 1-chains, or cycles, is therefore
\begin{equation}
Z_1 = \ker(B) \subset \mathbb{Z}_2^E,
\end{equation}
and is called the cycle space of the lattice. 
The number of independent cycles (cf. Theorem 1.9.5 of~\cite{diestel2017graph}) is given by
\begin{equation}
\dim Z_1 = E - V + 1.
\label{eq:dimZ1}
\end{equation}

\begin{remark}
\label{rem:cycles}
Not all cycles are topologically distinct. Cycles that can be continuously deformed into one another without crossing holes of the surface must be identified. On the lattice, such local deformations correspond to adding or removing the boundary of a face, see Fig. \ref{fig_cycle_plaquette}.
\end{remark}
\begin{figure}[h!]
    \centering
    \includegraphics[width=0.5\linewidth]{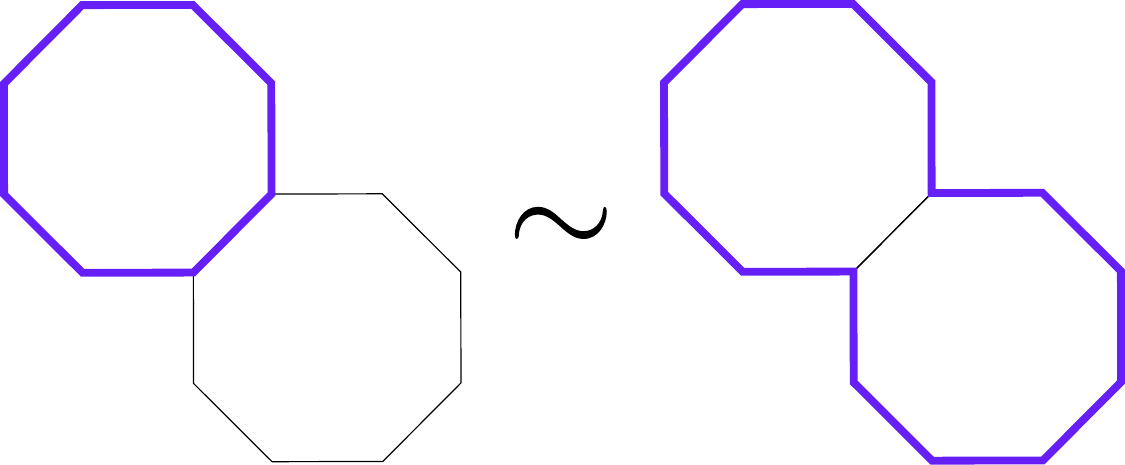}
    \caption{Two cycles (blue) that differ by a face boundary are considered equivalent.}
    \label{fig_cycle_plaquette}
\end{figure}
\begin{definition}
    A boundary is any cycle which is a linear combination of face boundaries.
\end{definition}
Each face $f\in F(\textbf{G})$ defines a boundary cycle $\partial f \in \mathbb{Z}_2^E$, given by the set of edges surrounding that face. 
\begin{definition} The face-edge incidence matrix
\begin{align}
C \in \mathbb{Z}_2^{F \times E},\qquad
C_{f\,e} =
\begin{cases}
1 & \text{if $e \in \partial f$},\\
0 & \text{otherwise}.
\end{cases}
\end{align}
\end{definition}
Given that we are working with a cellulation of a smooth surface, the condition $BC^T=0$ is satisfied. The subspace of boundaries is then given by the row space of $C$, $\text{Rowspace}(C)$, which is the vector space generated by all linear combinations of rows of $C$. Boundaries are precisely the cycles that can be shrunk to a point on the surface. On a closed genus-$g$ surface $\Sigma_g$, all edges belong to exactly 2 faces, so the sum (mod 2) of all face boundaries is trivial. This means only $F-1$  rows of $C$ are independent and therefore $\text{rank}(C)=F-1$.

\begin{definition}
The first homology group is defined as the quotient
\begin{equation}
\mathrm{H}_1(\Sigma_g,\mathbb{Z}_2)
=
Z_1 \big/ \text{Rowspace}(C),
\end{equation}
which means the collection of cycles up to boundary deformations. Two cycles are therefore considered equivalent if they differ by a sum of face boundaries, cf. \textit{Remark}~\ref{rem:cycles}. 
\end{definition}
This agrees with the general topological result
\begin{equation}
\dim \mathrm{H}_1(\Sigma_g,\mathbb{Z}_2) = 2g.
\label{eq:H_1}
\end{equation}
On a genus-$g$ surface there are thus $2g$ such independent cycles, which we denote
\begin{equation}
\{\gamma_1,\gamma_2,\dots,\gamma_{2g-1},\gamma_{2g}\},
\label{eq:wilson_loops}
\end{equation}
Wilson loop operators, relevant for describing the Hilbert space of the Kitaev model on such lattices, have support on these non-trivial cycles.

\begin{example}
To complete the characterization of T2.1, we evaluate cycle and homology dimensions: $\dim Z_1=E-V+1=9$, $\dim \mathrm{H}_1=\dim Z_1-(F-1)=9-5=4=2g$.
Similarly, for T4.1 we obtain $\dim Z_1=E-V+1=13$ and $\dim \mathrm{H}_1=\dim Z_1-(F-1)=13-5=8=2g$.
\label{ex:wilson_loops}
\end{example}

\section{Three-colored Kitaev model}
\label{sec:kitaev_model}

Having defined the finite PBC lattice and its topology, we now introduce the quantum model we study. We refer to the seminal paper by Kitaev and following publications \cite{Kitaev2006Anyons,Yao2009AlgebraicSpinLiquid,Lieb1994FluxPhase,Vidal2008PerturbativeKitaev,Cassella_2023,Lenggenhager2025HyperbolicSpinLiquids,Pedrocchi2011PhysicalSolutionsKitaev,ZschockeVojta2015FiniteSizeKitaev} for in-depth analysis and details.

We consider spin-$\frac12$ degrees of freedom on the vertices of a trivalent lattice, where each edge is assigned one of three types, conventionally labeled by $\alpha \in \{x,y,z\}$, such that at every vertex the three incident edges carry different labels. In the absence of the three-spin interaction term, a scenario considered throughout this work, the Kitaev Hamiltonian is defined as
\begin{equation}
 H = -\sum_{\langle ij\rangle_\alpha}
J_\alpha\, \sigma_i^\alpha  \sigma_j^\alpha,
\label{eq:H_Kitaev}
\end{equation}
where $\langle ij\rangle_\alpha$ denotes a nearest-neighbor connection of type $\alpha$, and $\sigma_i^\alpha$ are Pauli operators acting on site $i$.

\subsection{Edge coloring from the group construction}

In our construction, the lattice arises from the permutation action of the finite quotient $Q$ on oriented flags. Since the quotient lattice is bipartite and trivalent, Kőnig's line-coloring theorem guarantees a proper three-edge-coloring  (see Prop. 5.3.1, p.127 of \cite{diestel2017graph}). The bonds can therefore be colored by $\alpha\in\{x,y,z\}$ so that exactly one bond of each type meets at every site, defining a consistent three-colored Kitaev model.

\subsection{Majorana representation and link operators}
In order to remain compatible with the established terminology in physics literature, from now on we denote vertices, edges and faces as sites, links and plaquettes respectively. Following Kitaev, we represent spins using four Majorana fermions per site. We have matter fermions $c_i$ and gauge fermions $b_i$, satisfying
\begin{equation}
    \{c_i,c_j\}=2\delta_{ij},\qquad
\{b_i^\alpha,b_j^\beta\}
=2\delta_{ij}\delta_{\alpha\beta},\qquad
\{c_i,b_j^\alpha\}=0.
\end{equation}
The spin-to-fermion mapping is
\begin{equation}
 \sigma_i^\alpha = i\,  b_i^\alpha  c_i.
 \label{eq:kitaev_mapping}
\end{equation}

On a bond $(i,j)$ of type $\alpha$, we define the link operator
\begin{equation}
{u}_{ij} = i\,  b_i^\alpha  b_j^\alpha,
\end{equation}

The bond term becomes $\sigma_i^\alpha \sigma_j^\alpha = -i\,{u}_{ij}\, c_i  c_j$ and the operators ${u}_{ij}$ satisfy
\begin{align}
{u}_{ij} &= -{u}_{ji}, \nonumber\\
{u}_{ij}^2 &= 1, \nonumber\\
[{u}_{ij},H] &= 0.
\end{align}

Since all ${u}_{ij}$ commute with the Hamiltonian and with each other, they can be simultaneously diagonalized. The Hilbert space thus decomposes into sectors labeled by their eigenvalues $u_{ij} = \pm 1$ and the Hamiltonian reduces to a quadratic form in the matter Majorana fermions. We will give details about this in the next subsection. The resulting quadratic form is:
\begin{equation}
H = \sum_{\langle ij\rangle_\alpha}
J_\alpha\, i\,u_{ij}\,c_i c_j.
\label{eq:H_Kitaev_Majorana}
\end{equation}
\subsection{Flux operators, global constraints and superselection sectors}
Let $p$ be a face with cyclically ordered boundary vertices $(i_1,i_2,\dots,i_{|p|})$, with $i_{|p|+1}\equiv i_1$. At every vertex $i_k$, the two edges of $\partial p$ incident on $i_k$ carry two different colors; we denote by $\beta_k\in\{x,y,z\}$ the unique third color. The plaquette operator can then be written in spin variables as
\begin{equation}
W_p=\prod_{k=1}^{|p|}\sigma_{i_k}^{\beta_k},
\label{eq:plaquette_spin}
\end{equation}
Using Eq.~(\ref{eq:kitaev_mapping}), the matter fermions $c_{i_k}$ occur pairwise and cancel along the closed loop, leaving a product of link operators,
\begin{equation}
W_p=\prod_{k=1}^{|p|}u_{i_k i_{k+1}} .
\label{eq:plaquette_u}
\end{equation}
The cyclic ordering fixes the orientation with which each link variable enters the product, since $u_{ij}=-u_{ji}$.

The plaquette operators are gauge invariant: a local gauge transformation at a vertex reverses the signs of the two boundary links incident on that vertex, leaving their product unchanged. Moreover, since every link operator is conserved, $[u_{ij},H]=0$, it follows that
\begin{equation}
[W_p,H]=0,
\qquad
[W_p,W_{p'}]=0,
\qquad
W_p^2=1.
\end{equation}
Hence every plaquette flux $W_p=\pm1$ is a conserved quantity. 
On a closed surface the plaquette fluxes are not independent. Since every edge belongs to exactly two faces, the product over all plaquettes satisfies, with our orientation convention,
\begin{equation}
\prod_{p}W_p=1.
\label{eq:flux_condition}
\end{equation}
Consequently, only $F-1$ of the $F$ plaquette fluxes are independent, giving $2^{F-1}$ allowed plaquette flux configurations. Any change of flux configuration must flip an even number of plaquette eigenvalues.

Plaquette fluxes, however, do not completely specify the gauge sector on a surface of genus $g$. The additional degrees of freedom are characterized by $2g$ independent Wilson loop eigenvalues, giving $2^{2g}$ topological sectors, cf. Eqs.~(\ref{eq:H_1}), (\ref{eq:wilson_loops}).  

We therefore label a flux sector by
\begin{equation}
\bar\phi=(W_{p_1},\dots,W_{p_F}),
\qquad
W_{p_i}=\pm1,
\qquad
\prod_p W_p=1,
\label{eq:signature_flux}
\end{equation}
and a topological sector by
\begin{equation}
\bar\omega=(W_{\gamma_1},\dots,W_{\gamma_{2g}}),
\qquad
W_{\gamma_i}=\pm1.
\label{eq:signature_wilson}
\end{equation}
Thus there are $2^{F-1}$ flux sectors and $2^{2g}$ topological sectors, for a total of $2^{F-1+2g}$ gauge-invariant sectors.

The combined signature
\begin{equation}
s=(\bar\omega,\bar\phi)
\end{equation}
specifies an equivalence class of link configurations given by a certain gauge configuration. Choosing a representative $\{u_{ij}\}^{(s)}$ from this class fixes the quadratic Majorana Hamiltonian in that sector. The Hilbert space consequently decomposes in so-called superselection sectors
\begin{equation}
\mathcal H\simeq\bigoplus_s\mathcal H^{(s)},
\end{equation}
and within each sector the Kitaev model reduces to a free-fermion problem in the fixed $\mathbb{Z}_2$ gauge background $\{u_{ij}\}^{(s)}$.
\subsection{Majorana Hamiltonian}
Once the adjacency matrix $\mathrm{Adj}_{ij}$, the bond's colorings $\alpha$ and the gauge background $\{u_{ij}\}^{(s)}$ are fixed, we have all the necessary ingredients to construct the quadratic Majorana Hamiltonian

\begin{equation}
H^{(s)} = \frac{i}{4} \sum_{i,j} A^{(s)}_{ij}\, c_i c_j,
\label{eq:H_quadratic}
\end{equation}

where $A^{(s)}$ is a real antisymmetric matrix, $A^{(s)}_{ij} = -A^{(s)}_{ji}$. For a nearest-neighbour bond $\langle ij\rangle$ (this information is contained in $\mathrm{Adj}_{ij}$) of color $\alpha$, the corresponding matrix element is
\begin{equation}
A^{(s)}_{ij} = 2 J_\alpha\, u_{ij}^{(s)},
\qquad
A^{(s)}_{ji} =  -2 J_\alpha\, u_{ij}^{(s)},
\end{equation}
while $A^{(s)}_{ij}=0$ otherwise. Since $A^{(s)}$ is real and antisymmetric, there exists an orthogonal matrix
$O^{(s)}$ such that
\begin{equation}
\bigl(O^{(s)}\bigr)^{\mathsf T}A^{(s)}O^{(s)}
=
\bigoplus_{m=1}^{V/2}
\begin{pmatrix}
0 & \varepsilon_m^{(s)}\\
-\varepsilon_m^{(s)} & 0
\end{pmatrix}.
\end{equation}
The transformed operators
$\boldsymbol{\gamma}=\bigl(O^{(s)}\bigr)^{\mathsf T}\mathbf c$,
where $\mathbf c=(c_1,\ldots,c_V)^{\mathsf T}$, are again Majorana fermions.
Pairing them defines $V/2$ canonical complex fermionic modes,
\begin{align}
a_m&=\frac{\gamma_{2m-1}+i\gamma_{2m}}{2},
\nonumber
\\
\{a_m,a_n^\dagger\}&=\delta_{mn},\quad
\{a_m,a_n\}=0.
\end{align}
The Hamiltonian then takes the diagonal form
\begin{equation}
H^{(s)}=
\sum_{m=1}^{V/2}
\varepsilon_m^{(s)}
\left(
a_m^\dagger a_m-\frac{1}{2}
\right),
\qquad
\varepsilon_m^{(s)}\geq 0,
\end{equation}
the energy of a many-body state with occupation numbers
$n_m\in\{0,1\}$ is
\begin{equation}
E^{(s)}(\{n_m\})=
-\frac{1}{2}\sum_{m=1}^{V/2}\varepsilon_m^{(s)}+
\sum_{m=1}^{V/2}n_m\varepsilon_m^{(s)}.
\end{equation}
The total occupation-number operator is $N_a=\sum_m n_m$ and the vacuum energy of the $(s)$ sector is thus $E_{\rm vac}^{(s)}=-\frac{1}{2}\sum_m \varepsilon_m^{(s)}$.

\section{Parity projection}
\label{sec:parity_projection}

For a fixed superselection
sector $s=(\bar\omega,\bar\phi)$, the quadratic Majorana Hamiltonian Eq.~(\ref{eq:H_quadratic})
acts in the enlarged Majorana Hilbert space and must be projected onto the
physical subspace. 

Transforming the spin Hamiltonian Eq.~(\ref{eq:H_Kitaev}) to the Majorana representation (\ref{eq:kitaev_mapping}) enlarges the Hilbert space. To recover the physical ground state, we thus need to project on the physical space of (\ref{eq:H_Kitaev}). To this end, we follow the construction of Pedrocchi, Chesi and Loss \cite{Pedrocchi2011PhysicalSolutionsKitaev}. We define at each site the operator
\beq
D_j=b_j^x b_j^y b_j^z c_j,
\qquad
D_j^2=\mathds 1,
\eeq
such that the eigenvalues of $D_j$ are $\pm 1$. It follows that the physical states satisfy $D_j= \mathds 1$ \cite{Pedrocchi2011PhysicalSolutionsKitaev}.
The full projector is therefore
\begin{equation}
P=\prod_{j=1}^{V}\frac{1+D_j}{2}.
\label{eq:global_parity_projector}
\end{equation}
Combining the site constraints into a global parity condition gives
\begin{equation}
\eta_{\rm geom}\,
\det\bigl(O^{(s)}\bigr)\,
(-1)^{N_a}\,
\prod_{\langle ij\rangle}u_{ij}^{(s)}
=1,
\label{eq:parityprojectioncondition}
\end{equation}
Here $N_a=\sum_{m=1}^{V/2}a_m^\dagger a_m$ is the occupation number of the canonical complex fermions. The orthogonal matrix $O^{(s)}$ transforms the original matter Majoranas $c_i$ to the canonical Majorana basis that block-diagonalizes $A^{(s)}$, while $\eta_{\mathrm{geom}}\in\{\pm1\}$ is a fixed sign determined by the chosen ordering of the Majoranas and the orientation of the bonds in $\prod_{\langle ij\rangle}u_{ij}^{(s)}$. A derivation is given in App.~\ref{app:physical_parity}. We denote the left-hand side evaluated on the Majorana vacuum ($N_a=0$) by
\begin{equation}
D_{\rm vac}^{(s)}
:=
\eta_{\rm geom}\,
\det\bigl(O^{(s)}\bigr)\,
\prod_{\langle ij\rangle}u_{ij}^{(s)} .
\label{eq:parityprojectioncondition1}
\end{equation}
The physical parity condition in Eq.\ref{eq:parityprojectioncondition} therefore requires
$(-1)^{N_a}=D_{\rm vac}^{(s)}$. We define the lowest physical
energy in sector $s$ as
\begin{equation}
E_{\rm phys}(s)
=
\min_{\substack{\{n_m\}\in\{0,1\}^{V/2}\\
(-1)^{N_a}=D_{\rm vac}^{(s)}}}
E^{(s)}(\{n_m\}).
\label{eq:Ephys_sector}
\end{equation}
Denoting the smallest
single-particle energy by $\varepsilon_{\min}^{(s)}=\min_m\varepsilon_m^{(s)}$,
\begin{equation}
E_{\rm phys}(s)
=
\begin{cases}
E_{\rm vac}^{(s)}, & D_{\rm vac}^{(s)}=+1,\\
E_{\rm vac}^{(s)}+\varepsilon_{\min}^{(s)}, & D_{\rm vac}^{(s)}=-1.
\end{cases}
\end{equation}
The physical ground state energy of the finite cluster is then obtained by
minimizing over all superselection sectors,
\begin{equation}
E_{\rm GS}
=
\min_s E_{\rm phys}(s)
=
\min_{\bar\omega,\bar\phi}
E_{\rm phys}(\bar\omega,\bar\phi).
\label{eq:physical_GS_energy}
\end{equation}
We study the effect of parity projection on the resulting ground state energies in App.~\ref{app:projection_effects}.

\section{Numerical results}
\label{sec:Results}

Here we numerically solve the complete finite-size problem on compact hyperbolic quotients. For every quotient considered and every value of $J_x=J_y=J$, $J_z=1$, we enumerate all $2^{F-1+2g}$ gauge-invariant sectors, including both plaquette fluxes and the $2g$ non-contractible Wilson loops, and impose the physical parity condition separately in every sector. This allows us to determine the exact physical ground state of each finite PBC cluster and to study independently its energy density, its deviation from the homogeneous $\pi$-flux sector, the finite-size behavior of the flux gap, the vison gap and the symmetry origin of ground state flux degeneracies.

\subsection{Solved quotients catalogue}
The results are summarized in Table~\ref{tab:quotient_clusters}.
We group the finite PBC clusters according to the type $\{p,q\}$ of the underlying regular map $\mathcal M_Q$. The quotients $ {\rm T}g.n$ thus play the role of increasing system size compatible with the corresponding lattice type $\{p,q\}$ in analogy to increasing the system size on Euclidean lattices by adding lattice sites to the boundaries and compactifying on a torus.

\begin{table}[t]
\centering
\begin{tabular}{c|c|c|c|c|c|c}
$\{p,q\}$ & Quotient & $V$ & $E$ & $F$ & $g$ & $2^{F-1+2g}$ \\
\hline
\{8,3\}
& T2.1 & 16 & 24 & 6 & 2 & $2^9=512$ \\
& T3.2 & 32 & 48 & 12 & 3 & $2^{17}=131072$ \\
& T5.1 & 64 & 96 & 24 & 5 & $2^{33}\simeq 8.59\times10^9$ \\
\hline
\{12,3\}
& T3.3 & 16 & 24 & 4 & 3 & $2^9=512$ \\
& T4.1 & 24 & 36 & 6 & 4 & $2^{13}=8192$ \\
& T7.2 & 48 & 72 & 12 & 7 & $2^{25}\simeq 3.36\times10^7$ \\
& T9.1 & 64 & 96 & 16 & 9 & $2^{33}\simeq 8.59\times10^9$ \\
& T10.3 & 72 & 108 & 18 & 10 & $2^{37}\simeq 1.37\times10^{11}$ \\
\hline
\end{tabular}
\caption{Finite PBC clusters studied numerically, grouped according to the Schläfli type $\{p,q\}$ of the underlying map.}
\label{tab:quotient_clusters}
\end{table}
For each quotient, we construct the corresponding finite Kitaev Hamiltonian and exactly enumerate all superselection sectors. In each sector, the Majorana Hamiltonian is diagonalized and the physical-state projection is imposed, allowing us to determine the physical ground state energy and the corresponding ground state gauge sector. We repeat this procedure as a function of the anisotropy parameter $J_x=J_y\equiv J$, keeping $J_z=1$. 

\subsection{Ground state energy density}
As a first comparison between different finite quotients and lattice families, in Fig. \ref{fig_egs_density_8_3},\ref{fig_egs_density_12_3} we show the resulting ground state energy density $E_{\mathrm{GS}}/V$ as a function of $J$ for two distinct $\{p,q\}$ geometries.
\begin{figure}[h!]
    \centering
    \includegraphics[width=0.95\linewidth]{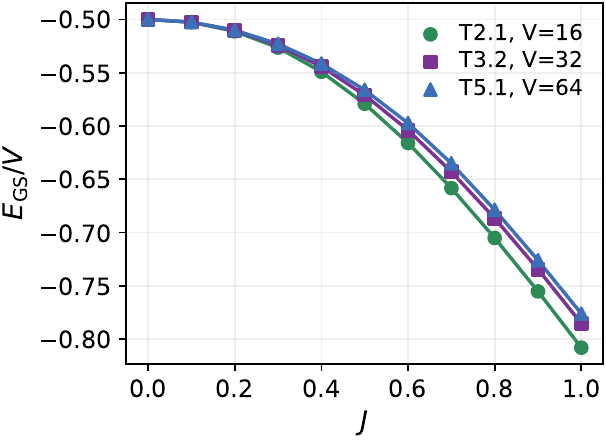}
    \caption{$\{8,3\}$ family ground state energy density}
    \label{fig_egs_density_8_3}
\end{figure}
\begin{figure}[h!]
    \centering
    \includegraphics[width=0.95\linewidth]{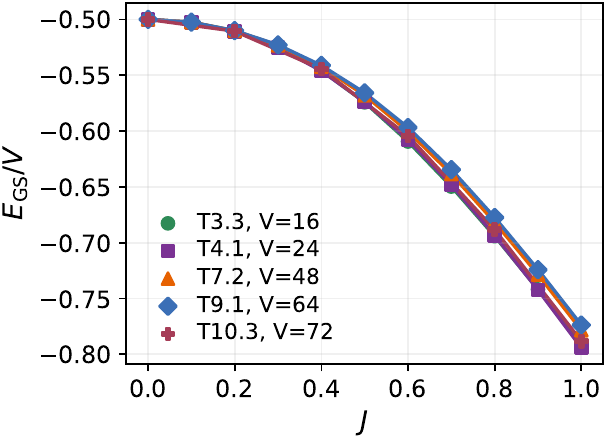}
    \caption{$\{12,3\}$ family ground state energy density}
    \label{fig_egs_density_12_3}
\end{figure}
The ground state energy densities at the strongly anisotropic point $J_x=J_y=0$, $J_z=1$ are consistent with the fully dimerized limit of the toric code phase. In this limit, only the $z-$bond couplings contribute to the Hamiltonian, and the lattice decomposes into $V/2$ independent dimers. Each dimer contributes an energy -1, giving $E_{\mathrm{GS}}=-V/2$ and therefore $E_{\mathrm{GS}}/V=-1/2$, independently of the quotient. The ground state is also extensively degenerate: each dimer has two energetically equivalent aligned spin configurations, so that the total degeneracy is $2^{V/2}$. This large degeneracy is lifted as soon as $J_x$ and $J_y$ become nonzero.

\subsection{Deviation from homogeneous $\pi$-flux configuration}
We now analyze what sector is actually selected for the ground state. We do this by plotting the energies of the physical ground state $E_{\mathrm{GS}}$ and of the lowest-energy state within the homogeneous $\pi$-flux sector. We define
\begin{equation}
\bar\phi_{\pi}=(-1,\ldots,-1),
\qquad
E_{\pi}=
\min_{\bar\omega}
E_{\mathrm{phys}}\!\left(s=(\bar\omega,\bar\phi_{\pi})\right).
\end{equation}
where the minimization is taken over all Wilson sectors $\bar\omega\in\{\pm1\}^{2g}$ compatible with the homogeneous plaquette-flux pattern $W_p=-1\;\forall\,p$. We then compare $E_{\pi}$ with $E_{\mathrm{GS}}$, by defining the corresponding energy difference $\Delta E_{\pi}/V=(E_{\pi}-E_{\mathrm{GS}})/V$. We notice how the value of this quantity across parameter space is strongly quotient-dependent: Fig.\ref{fig_DeltaEpi_density_8_3} shows a clear decrease of $\Delta E_{\pi}/V$ for $\{8,3\}$ quotients of increasing size, meaning that the energy of the homogeneous flux configuration becomes very close to the actual ground state energy, whereas Fig.\ref{fig_DeltaEpi_density_12_3} shows the opposite trend for the $\{12,3\}$ family, even though $\Delta E_{\pi}/V$ is one order of magnitude smaller in this case. Thus the $\pi$-flux configuration is not a priori the favored one for clusters of modest size. This is confirmed by the fraction of flipped plaquettes over the total number of plaquettes, $\eta=N_{\text{flipped}}/F$ in Fig.\ref{fig_nflipped-invV}.

These deviations from homogeneous $\pi$ flux do not represent a failure of Lieb's theorem, but rather reflect that its assumptions are not satisfied by these finite PBC Hamiltonians \cite{Lieb1994FluxPhase,Eschmann_2020_PRB}. Although the quotients considered here are reflexible before coloring them, Lieb's reflection-positivity argument requires a bond-cutting reflection that preserves the bond-colored Hamiltonian and divides the lattice into two disconnected halves exchanged by the reflection. None of the relevant reflections has this property for the quotients studied here. Moreover, the coloring based on perfect matchings we employ is not generally preserved by these reflections. Lieb's argument therefore does not determine their energy-minimizing flux sector. At the same time, the inapplicability of the theorem does not exclude a homogeneous $\pi$-flux ground state, as demonstrated by T3.3. Conversely, Examples~14 and~15 show that T7.2 and T5.1 select, respectively, homogeneous zero flux and nonuniform flux patterns. Reference~\cite{Lenggenhager2025HyperbolicSpinLiquids} instead constructs a reflection-compatible coloring for the infinite $\{8,3\}$ lattice and uses Lieb's argument to select homogeneous $\pi$ flux in the thermodynamic setting. Its large finite PBC calculation retains this flux background even though that particular quotient does not preserve all bond-cutting reflections. Our finite-sector minimization therefore complements, rather than contradicts, the thermodynamic result.


\begin{figure}
    \centering
    \includegraphics[width=0.95\linewidth]{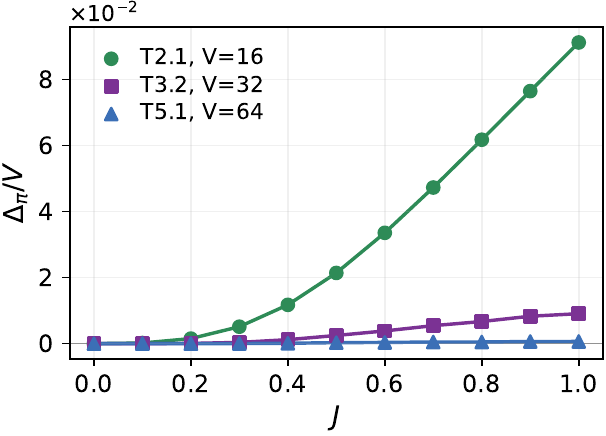}
    \caption{$\Delta E_{\pi}/V$ for the $\{8,3\}$ family as a function of $J$.}
    \label{fig_DeltaEpi_density_8_3}
\end{figure}
\begin{figure}
    \centering
    \includegraphics[width=0.95\linewidth]{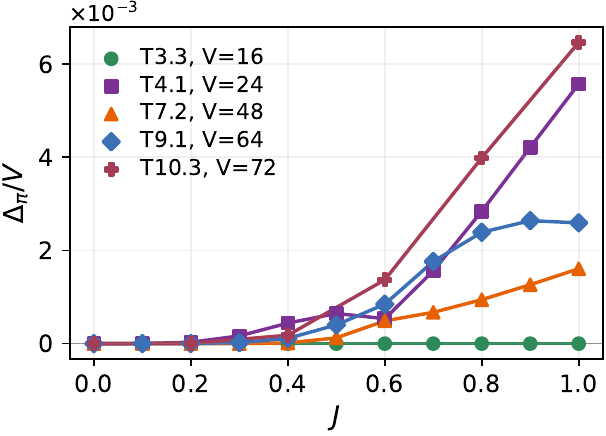}
    \caption{$\Delta E_{\pi}/V$ for the $\{12,3\}$ family as a function of $J$.}
    \label{fig_DeltaEpi_density_12_3}
\end{figure}
\begin{figure}[t]
    \centering
    \includegraphics[width=\columnwidth]{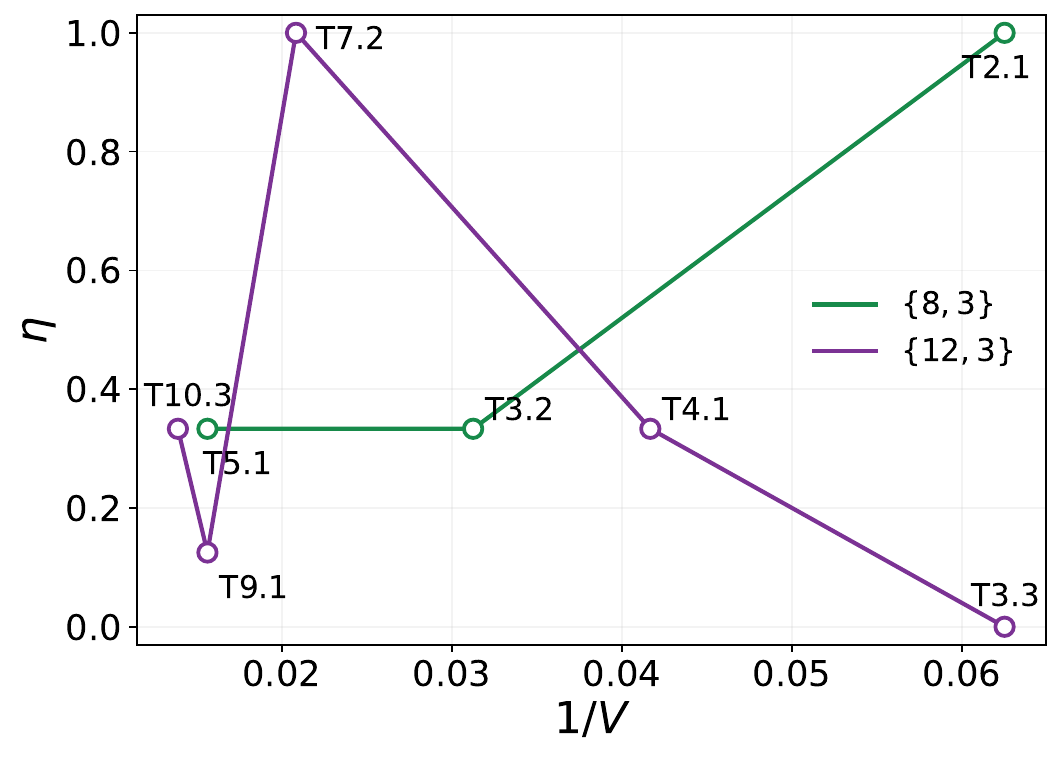}
    \caption{
    Fraction of plaquettes whose flux is flipped relative to the uniform $\pi$-flux configuration, $\eta=N_{\text{flipped}}/F$, in the exact projected ground state at the isotropic point $J=1$, plotted against inverse system size $1/V$, for the two families of quotients.}
    \label{fig_nflipped-invV}
\end{figure}
\subsection{Ground state sector degeneracy and lattice symmetries}

The analysis above determines which plaquette flux configurations are energetically favored. We now ask a different question: how many distinct superselection sectors realize the same ground state energy?
We recall that two sectors can differ locally, through their plaquette flux pattern, globally, through their Wilson loop eigenvalues, or through both.

We restrict the following discussion to $J>0$. At $J=0$ the decoupled dimer limit has an additional extensive many body degeneracy which is not the sector degeneracy considered here.

We define the set of ground state sectors by
\begin{equation}
\mathcal S_{\mathrm{GS}}(J)
=
\left\{
(\bar\omega,\bar\phi):
E_{\mathrm{phys}}(\bar\omega,\bar\phi;J)
=
E_{\mathrm{GS}}(J)
\right\},
\label{eq:S_GS}
\end{equation}
and its cardinality by
\begin{equation}
d_{\mathrm{GS}}(J)
=
|\mathcal S_{\mathrm{GS}}(J)|.
\label{eq:GSsectordeg}
\end{equation}
Here $d_{\mathrm{GS}}$ counts distinct ground state superselection sectors and should not be confused with a possible additional many body degeneracy of the matter Hamiltonian within a fixed superselection sector.

It is useful to resolve this degeneracy according to the two labels $(\bar\omega,\bar\phi)$. For a fixed plaquette flux configuration we define
\begin{equation}
E_{\mathrm{flux}}(\bar\phi;J)
=
\min_{\bar\omega}
E_{\mathrm{phys}}(\bar\omega,\bar\phi;J),
\label{eq:Eflux}
\end{equation}
and hence the set of plaquette flux configurations appearing in the ground state sector manifold,
\begin{equation}
\Phi_{\mathrm{GS}}(J)
=
\left\{
\bar\phi:
E_{\mathrm{flux}}(\bar\phi;J)
=
E_{\mathrm{GS}}(J)
\right\}.
\label{eq:GSfluxmanifold}
\end{equation}
We denote their number by
\begin{equation}
d_{\bar \phi}(J)
=
|\Phi_{\mathrm{GS}}(J)|.
\end{equation}

For each $\bar\phi\in\Phi_{\mathrm{GS}}$, we may then ask how many Wilson sectors remain at the ground energy while keeping that particular plaquette flux configuration fixed. We define
\begin{equation}
\Omega_{\mathrm{GS}}(\bar\phi;J)
=
\left\{
\bar\omega:
E_{\mathrm{phys}}(\bar\omega,\bar\phi;J)
=
E_{\mathrm{GS}}(J)
\right\},
\label{eq:Omega_GS}
\end{equation}
and
\begin{equation}
d_{\bar \omega}(\bar\phi;J)
=
|\Omega_{\mathrm{GS}}(\bar\phi;J)|.
\end{equation}
Thus, $d_{\phi}$ counts how many distinct ground state plaquette flux configurations exist, whereas $d_{\omega}(\bar\phi;J)$ counts, for one particular ground state flux configuration $\bar\phi$, how many distinct Wilson loop configurations can accompany it without changing the ground state energy.

The total ground state sector degeneracy is obtained by adding these Wilson multiplicities over all ground state flux configurations,
\begin{equation}
d_{\mathrm{GS}}(J)
=
\sum_{\bar\phi\in\Phi_{\mathrm{GS}}(J)}
d_{\bar \omega}(\bar\phi;J).
\label{eq:GSdeg_general}
\end{equation}

More generally, the values $d_{\omega}(\bar\phi;J)$ need not a priori be the same for different $\bar\phi$. If they are all equal, we may denote their common value simply by $d_{\omega}(J)$, in which case Eq.~(\ref{eq:GSdeg_general}) becomes
\begin{equation}
d_{\mathrm{GS}}(J)
=
d_{\bar \phi}(J)\,d_{\bar \omega}(J).
\label{eq:GSdeg_decomposition}
\end{equation}
This may not hold in general, but it does hold for all quotients considered in this analysis.
The observed degeneracies have a direct connection to the symmetries of the Kitaev Hamiltonian. The relevant symmetry group is not simply the automorphism group of the underlying graph, because the Kitaev Hamiltonian distinguishes the three bond types through the spin components appearing on them. A permutation of the lattice sites may therefore preserve the graph while failing to preserve the Hamiltonian.

To see the constraint imposed by the Hamiltonian, consider an $\alpha$-type bond $\langle ij\rangle_\alpha$ and its two-body term
\begin{equation}
J_\alpha\,\sigma_i^\alpha\sigma_j^\alpha.
\label{eq:bond_alpha}
\end{equation}

Let $\pi$ be a lattice automorphism mapping the $\alpha$-type bond $\langle ij\rangle_\alpha$ onto the $\beta$-type bond$\langle \pi(i)\pi(j)\rangle_\beta$. Associate with every site $k$ a signed permutation matrix $M_k$ acting on its spin axes. Preservation of the bond interaction requires the transformations at $i$ and $j$ to send the $\alpha$ axis to the $\beta$ axis with the same sign, together with $J_\alpha=J_\beta$. These onsite transformations must be realizable as proper spin rotations rather than arbitrary orthogonal transformations. We therefore require $M_k\in\mathrm{SO}(3)$ at every site $k$ ($\det M_k=1$); orthogonal transformations with determinant $-1$, corresponding to improper rotations or reflections, are excluded. 
We define $\mathcal G_{\mathrm{phys}}(J)$ as the subgroup of lattice automorphisms $\pi$ for which onsite transformations satisfying this condition exist. These conditions are then imposed simultaneously on every edge of the lattice. This construction also explains the dependence of $\mathcal G_{\mathrm{phys}}$ on $J$. Along $J_x=J_y=J$, $J_z=1$ considered here, the $x$ and $y$ bonds may be interchanged because they carry equal couplings, whereas for $J\neq1$ the $z$ bonds cannot be interchanged with them. At the isotropic point $J=1$, all three couplings are equal, so symmetries involving permutations of all three bond types can also be admitted.

To make the statement concrete, we start from one particular ground state sector $s_{\mathrm{GS}}$. Applying every element of $\mathcal G_{\mathrm{phys}}$ produces its symmetry orbit,
\begin{equation}
\operatorname{Orb}_{\mathcal G_{\mathrm{phys}}}(s_{\mathrm{GS}})
=
\left\{
g\cdot s_{\mathrm{GS}}:
g\in \mathcal G_{\mathrm{phys}}
\right\}.
\label{eq:Orbit}
\end{equation}
Every sector in this orbit has exactly the same energy, because the transformations are symmetries of the Hamiltonian. Some transformations, however, may leave the original sector unchanged. These transformations form the stabilizer
\begin{equation}
\operatorname{Stab}_{\mathcal G_{\mathrm{phys}}}(s_{\mathrm{GS}})
=
\left\{
g\in \mathcal G_{\mathrm{phys}}:
g\cdot s_{\mathrm{GS}}=s_{\mathrm{GS}}
\right\}.
\label{eq:Stabilizer}
\end{equation}
The orbit-stabilizer theorem \cite{Rotman1995TheoryOfGroups} then gives
\begin{equation}
\left|
\operatorname{Orb}_{\mathcal G_{\mathrm{phys}}}(s_{\mathrm{GS}})
\right|
=
\frac{|\mathcal G_{\mathrm{phys}}|}
{|\operatorname{Stab}_{\mathcal G_{\mathrm{phys}}}(s_{\mathrm{GS}})|}.
\label{eq:GSorbit}
\end{equation}

A highly symmetric ground state sector has a large stabilizer and therefore a small orbit. Conversely, if only a small subgroup leaves the sector unchanged, many distinct symmetry-related ground state sectors are generated.

We explicitly construct the action of $\mathcal G_{\mathrm{phys}}(J)$ on the complete sector labels $(\bar\omega,\bar\phi)$ for every quotient included in the symmetry analysis and every sampled value $J>0$. In all cases, the complete set $\mathcal S_{\mathrm{GS}}(J)$ of ground-state sectors forms a single symmetry orbit. Its degeneracy therefore follows directly from the orbit--stabilizer relation in Eq.~(\ref{eq:GSorbit}). The evolution of $d_{\bar\phi}$, $d_{\bar\omega}$, and $d_{\mathrm{GS}}$ with $J$ is reported in App.~\ref{app:GS_degeneracy_J}. At the isotropic point, the corresponding values of $|\mathcal G_{\mathrm{phys}}|$, $d_{\bar\phi}$, $d_{\bar\omega}$, $d_{\mathrm{GS}}$, and the stabilizer size are summarized in Table~\ref{tab:GS_symmetry_J1}.
\begin{figure}[h!]
    \centering
    \includegraphics[width=\linewidth]{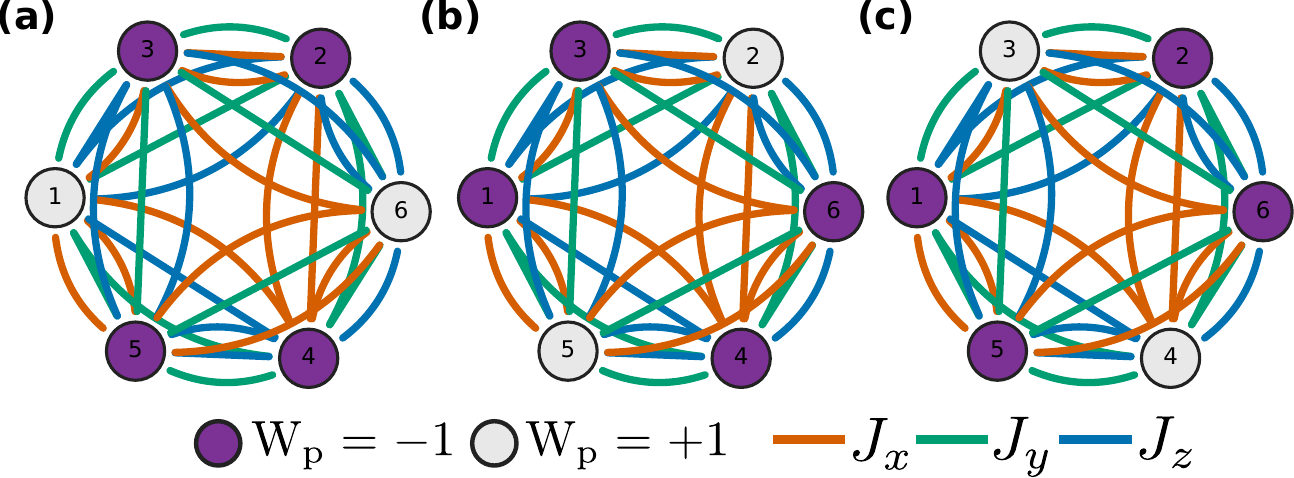}
    \caption{The three ground state plaquette flux configurations of T4.1 at $J=1$. In this dual graph representation, a circle represents a plaquette, with purple and white corresponding to $W_p=-1$ and $W_p=+1$, respectively. The colored curves represent the physical edges shared by pairs of plaquettes, with colors indicating the couplings $J_x$, $J_y$, and $J_z$. The three configurations are related by physical symmetries of the Kitaev Hamiltonian.}
    \label{fig_T41GSfluxfig}
\end{figure}

\begin{table}[h!]
\centering
\begin{tabular}{c|r|r|r|r|r}
\hline
Quotient & $|\mathcal G_{\mathrm{phys}}|$ & $d_{\bar\phi}$ & $d_{\bar\omega}$ & $d_{\mathrm{GS}}$ & $|\mathrm{Stab}|$ \\
\hline
T2.1 & 96  & 1  & 1 & 1   & 96 \\
T3.2 & 192 & 3  & 1 & 3   & 64 \\
T3.3 & 96  & 1  & 6 & 6   & 16 \\
T4.1 & 144 & 3  & 1 & 3   & 48 \\
T5.1 & 384 & 3  & 2 & 6   & 64 \\
T7.2 & 288 & 1  & 3 & 3   & 96 \\
T9.1 & 384 & 48 & 4 & 192 & 2 \\
T10.3 & 432 & 3 & 1 & 3 & 144 \\
\hline
\end{tabular}
\caption{Ground state symmetry data at the isotropic point $J=1$. The stabilizer refers to a representative ground state sector $s_{\mathrm{GS}}$. In every case,
$d_{\mathrm{GS}}=|\mathcal G_{\mathrm{phys}}|/|\mathrm{Stab}|=d_{\bar\phi} d_{\bar\omega}$.}
\label{tab:GS_symmetry_J1}
\end{table}
\example{The T4.1 quotient at $J=1$ gives the simplest example of a degeneracy carried entirely by the plaquette flux labels. We find
\begin{equation}
d_{\mathrm{GS}}=3,
\qquad
d_{\phi}=3,
\qquad
d_{\omega}=1.
\end{equation}
There are therefore three distinct ground state sectors, shown in Fig.~\ref{fig_T41GSfluxfig}, and each has a different plaquette flux configuration. Once $\bar\phi$ is fixed, however, there is only one Wilson sector at the ground state energy.
The symmetry interpretation is particularly transparent. The physical symmetry group of T4.1 at $J=1$ has $144$ elements, while $48$ of them leave a chosen ground state sector unchanged. Equation~(\ref{eq:GSorbit}) therefore gives $d_{\mathrm{GS}}=\frac{144}{48}=3.$
Starting from any of the three panels in Fig.\ref{fig_T41GSfluxfig}, physical lattice symmetries generate the other two. The three flux patterns are therefore not three unrelated minima: they are three symmetry-related realizations of the same ground energy.

This also gives an operational way of distinguishing the three sectors. Suppose an experimental realization prepares the T4.1 cluster in its ground state manifold. For each plaquette $p$ the flux $W_p$ is a physical spin observable. If the boundary of a T4.1 plaquette is $(i_1,\ldots,i_{12})$, then
\begin{equation}
W_p
=
\prod_{k=1}^{12}
\sigma_{i_k}^{\beta_k},
\end{equation}
where $\beta_k\in\{x,y,z\}$ is the color of the unique bond incident on $i_k$ that does not belong to the boundary of $p$.

Measuring the six plaquette operators determines $\bar\phi$.
In our convention, $W_p=-1$ identifies a $\pi$-flux through plaquette $p$. The location of  $\pi$-fluxes is specified by three patterns in Fig.~\ref{fig_T41GSfluxfig}. Because $d_{\omega}=1$ for T4.1 at $J=1$, identifying $\bar\phi$ also identifies the complete ground state superselection sector.}
\example{For T7.2 at $J=1$,
\begin{equation}
d_{\mathrm{GS}}=3,
\qquad
d_{\bar\phi}=1,
\qquad
d_{\bar\omega}=3.
\end{equation}
The unique ground state plaquette flux configuration is the $0$-flux configuration
\begin{equation*}
\bar\phi=
(1,1,1,1,1,1,1,1,1,1,1,1),
\end{equation*}
while the three ground state Wilson configurations are
\begin{align*}
\bar\omega_1&=(1,1,-1,-1,1,-1,-1,-1,1,1,1,-1,1,-1),\\
\bar\omega_2&=(-1,-1,1,1,-1,-1,-1,-1,1,-1,1,-1,1,1),\\
\bar\omega_3&=(-1,-1,-1,-1,-1,1,1,1,1,-1,1,1,1,-1).
\end{align*}
Thus the three ground state sectors have identical local plaquette fluxes and differ only in their non-contractible Wilson loop eigenvalues.}

\example{For T5.1 at $J=1$,
\begin{equation*}
d_{\mathrm{GS}}=6,
\qquad
d_{\bar\phi}=3,
\qquad
d_{\bar\omega}=2.
\end{equation*}
The three ground state plaquette flux configurations are
\begin{align*}
\bar\phi_1={}&(1,-1,-1,-1,-1,1,1,-1,-1,-1,-1,1,\\
&\hspace{1.9em}-1,1,-1,1,-1,1,-1,-1,-1,-1,-1,1),\\
\bar\phi_2={}&(-1,1,-1,-1,1,-1,-1,-1,1,-1,1,-1,\\
&\hspace{1.9em}1,-1,1,-1,-1,-1,-1,-1,1,-1,1,-1),\\
\bar\phi_3={}&(-1,-1,1,1,-1,-1,-1,1,-1,1,-1,-1,\\
&\hspace{1.9em}-1,-1,-1,-1,1,-1,1,1,-1,1,-1,-1).
\end{align*}
For these three flux configurations, respectively, the set of ground state energy Wilson sectors has two elements, each specified by the $2g=10$ fluxes, cf. Eq.~(\ref{eq:Omega_GS})
\begin{align*}
\Omega_{\mathrm{GS}}(\bar\phi_1)
={}&\big\{
(1,-1,1,1,1,-1,1,1,-1,-1),\\
&\hspace{2.4em}(-1,1,1,1,-1,1,1,1,1,1)
\big\},\\
\Omega_{\mathrm{GS}}(\bar\phi_2)
={}&\big\{
(1,1,1,1,1,1,-1,-1,-1,1),\\
&\hspace{2.4em}(-1,1,-1,-1,1,1,1,1,1,1)
\big\},\\
\Omega_{\mathrm{GS}}(\bar\phi_3)
={}&\big\{
(1,1,-1,1,1,1,-1,1,1,-1),\\
&\hspace{2.4em}(1,-1,1,-1,-1,-1,1,-1,1,1)
\big\}.
\end{align*}
Thus the six ground state sectors consist of two Wilson sectors for each of the three plaquette flux configurations.}

These results are also important for defining flux excitations. If $d_{\phi}>1$, flipping an adjacent pair of plaquette fluxes can map one member of $\Phi_{\mathrm{GS}}$ into another and therefore have exactly zero energy cost without producing an excited state. We account explicitly for this possibility below when defining the vison gap.


\subsection{Vison gap scaling}

Having established the structure of the ground state manifold and the effect of physical projection, we now use the projected sector energies to study flux excitations. As shown above, the physical ground state need not occur in the homogeneous $\pi$ flux configuration and $\Phi_{\mathrm{GS}}$ may contain several distinct ground state flux configurations.

Because the plaquette fluxes satisfy the global constraint $\prod_p W_p=1$, a flux excitation on a closed quotient cannot be created by flipping a single plaquette. The minimal local modification instead flips an adjacent pair of plaquette eigenvalues. This is the simplest mechanism of having a vison excitation in a PBC cluster. For two adjacent plaquettes $p, q$, where adjacency is denoted as $p\sim q$, starting from a ground state flux configuration $\bar\phi\in\Phi_{\mathrm{GS}}$, we denote by $\bar\phi^{(p,q)}$ the configuration obtained through
\begin{equation*}
W_p\rightarrow-W_p,
\qquad
W_q\rightarrow-W_q,
\end{equation*}
while leaving all other plaquette eigenvalues unchanged. The corresponding physical excitation energy is
\begin{equation}
\Delta E(\bar\phi;p,q)
=E_{\mathrm{flux}}\!\left(\bar\phi^{(p,q)};J\right)-E_{\mathrm{GS}}(J).
\label{eq:two_vison_cost}
\end{equation}
When $\Phi_{\mathrm{GS}}$ contains more than one element, an adjacent plaquette-pair flip does not necessarily create an excitation. It can instead map one ground state flux configuration into another,
\begin{equation}
\bar\phi\in\Phi_{\mathrm{GS}},
\qquad
\bar\phi^{(p,q)}\in\Phi_{\mathrm{GS}},
\end{equation}
which effectively corresponds to a rotation in the ground state manifold and Eq.~(\ref{eq:two_vison_cost}) implies $\Delta E=0$. We then scan over all adjacent plaquette pairs starting from every $\bar\phi\in\Phi_{\mathrm{GS}}$, minimize the physical energy over all $\bar\omega$, and discard all targets that remain inside $\Phi_{\mathrm{GS}}$.

We define the finite-size vison gap as the smallest remaining positive adjacent two-flux excitation,
\begin{equation}
\Delta_{\mathrm v}(J)
=
\min_{\substack{
\bar\phi\in\Phi_{\mathrm{GS}}(J),\,p\sim q\\
\bar\phi^{(p,q)}\notin\Phi_{\mathrm{GS}}(J)
}}
\left[
E_{\mathrm{flux}}\!\left(\bar\phi^{(p,q)};J\right)
-
E_{\mathrm{GS}}(J)
\right].
\label{eq:visongap}
\end{equation}
Strictly speaking, $\Delta_{\mathrm v}$ is therefore a two-vison gap. This distinction becomes particularly important for quotients with a degenerate ground state flux manifold. For example, at $J=1$ the T9.1 quotient has $48$ distinct elements of $\Phi_{\mathrm{GS}}$. The lowest genuine positive adjacent two-vison excitation instead has $\Delta_{\mathrm v}\approx 0.133$.

Fig.~\ref{fig_visongapfig} shows the resulting vison gap at the isotropic point $J=1$ as a function of inverse system size for the $\{8,3\}$ and $\{12,3\}$ quotient families. The two families display markedly different finite-size scales. Within the $\{8,3\}$ family, $\Delta_{\mathrm v}$ decreases strongly between T2.1 and T3.2 and changes only weakly between T3.2 and T5.1. The $\{12,3\}$ family displays substantially smaller gaps, apart from the one for the largest T10.3 quotient, and a non-monotonic dependence on system size. 

\begin{figure}[h!]
    \centering
\includegraphics[width=0.95\columnwidth]{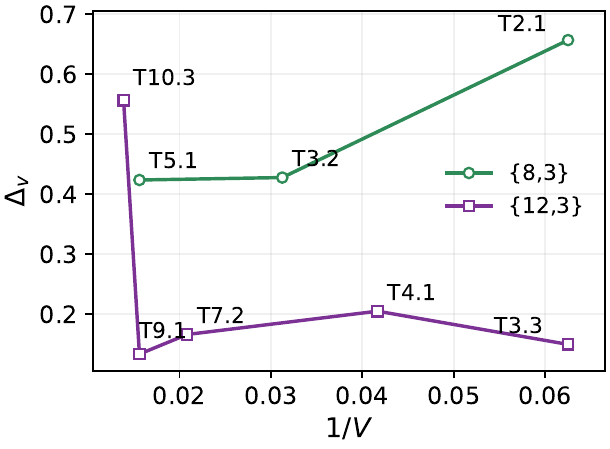}
    \caption{
    Finite-size dependence of the adjacent two-vison gap $\Delta_{\mathrm v}$ at the isotropic point $J=1$ for the $\{8,3\}$ and $\{12,3\}$ quotient families. For each quotient, the gap is obtained by considering every adjacent plaquette-pair flip relative to every ground state flux configuration $\bar\phi\in\Phi_{\mathrm{GS}}$, minimizing the physical energy over all Wilson sectors $\bar\omega$, and excluding flips whose target configuration remains inside $\Phi_{\mathrm{GS}}$. Lines are guides to the eye.
    }
    \label{fig_visongapfig}
\end{figure}

\subsection{Matter correlations at the isotropic point}
Finally, to further characterize the isotropic point, we compute the equal-time Majorana correlator in the exact physically projected ground state. For each graph distance $r$, we define
\begin{equation}
C(r)=\frac{1}{N_r}\sum_{d(i,j)=r}\left|\left\langle ic_i c_j\right\rangle
\right|,\end{equation}
where $N_r$ is the number of pairs at distance $r$. Figure~\ref{fig_matter-correlations} shows $C(r)$ for T5.1 and T10.3 at $J=1$, together with least-squares fits to the algebraic and exponential functions
\begin{equation*}
C_{\mathrm{alg}}(r)=A r^{-\alpha},
\qquad C_{\mathrm{exp}}(r)=B e^{-r/\xi},
\end{equation*}
where $A$ and $B$ are amplitudes, $\alpha$ is the algebraic decay exponent, and $\xi$ is the correlation length. The quality of each fit is quantified by the coefficient of determination $R^2$. We see that while the small number of available distances prevents an unambiguous determination of the asymptotic decay law, the shown results are compatible with the expected algebraic decay law at the isotropic point $J=1$.
\begin{figure}[h]
    \centering
    \includegraphics[width=0.95\columnwidth]{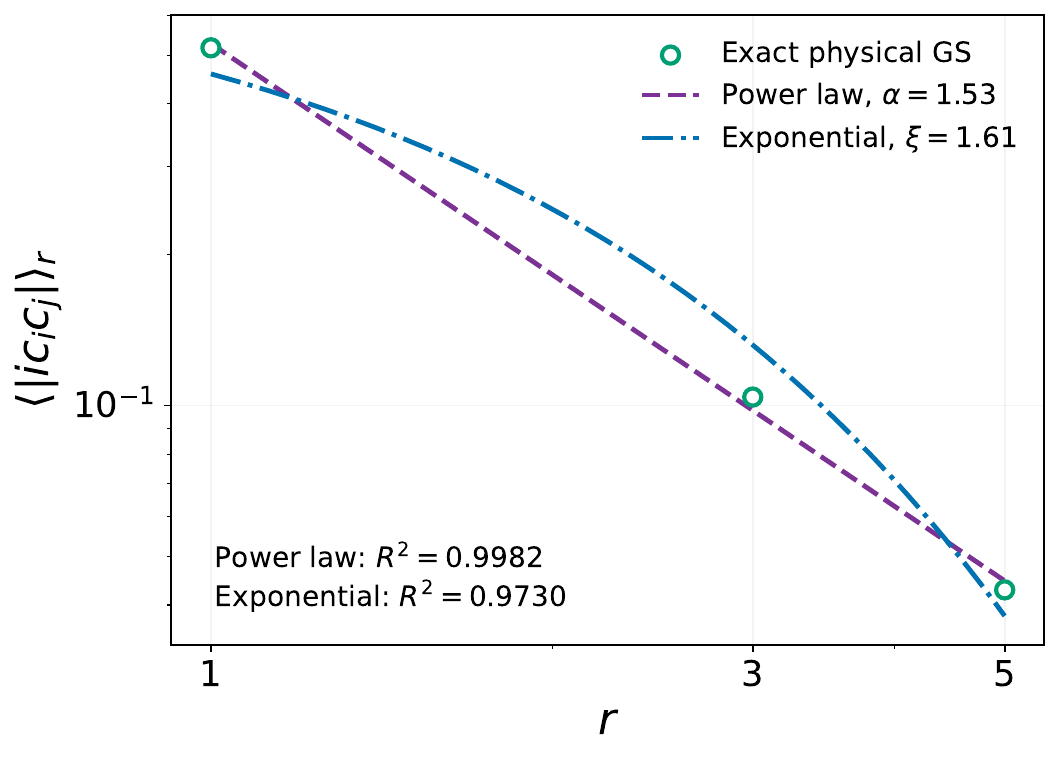}
    \vspace{0.1em}
    \includegraphics[width=0.95\columnwidth]{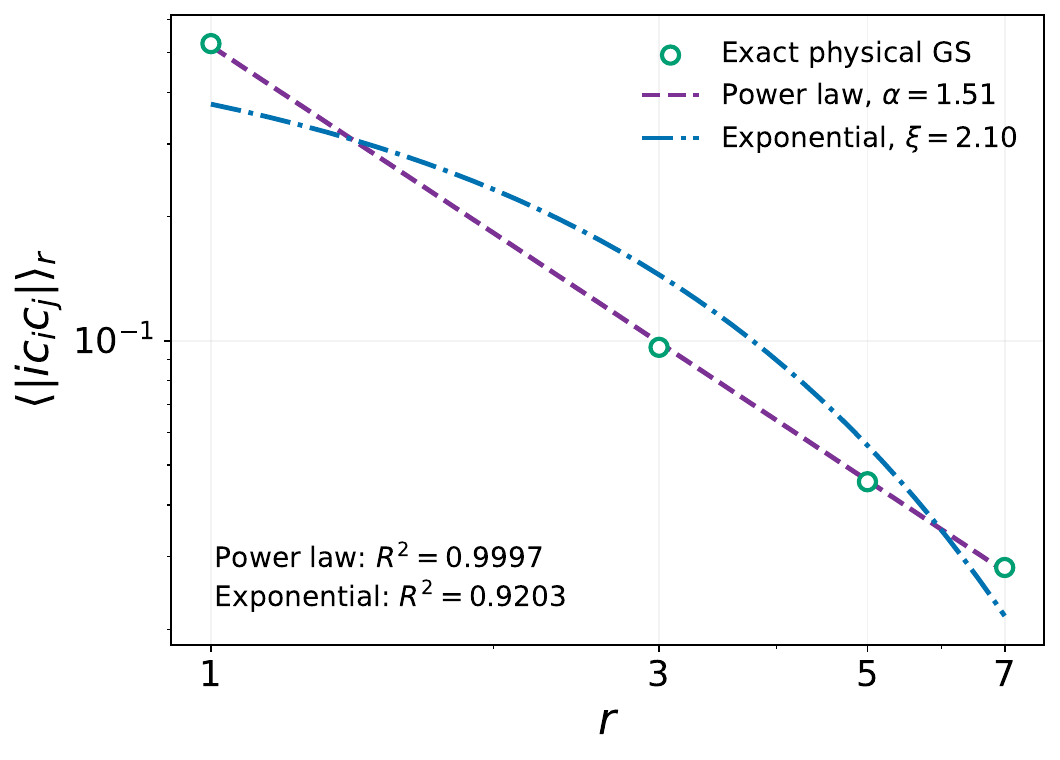}
    \caption{
    Distance-averaged equal-time Majorana correlations in the exact projected ground state at the isotropic point $J=1$ for T5.1 (top) and T10.3 (bottom). For T5.1, the best algebraic fit gives $\alpha\simeq1.53$ with $R^2\simeq0.998$, while the best exponential fit gives $\xi\simeq1.61$ with $R^2\simeq0.973$. For T10.3, using the accessible distances $r=1,3,5,7$, the algebraic fit gives $\alpha\simeq1.51$ with $R^2\simeq0.9997$, while the exponential fit gives $\xi\simeq2.10$ with $R^2\simeq0.9203$.
    }
    \label{fig_matter-correlations}
\end{figure}

\section{Outlook}
\label{sec:Conclusion}
%
Driven by the possibility of realizing spin Hamiltonians on current quantum hardware, in this work we have analyzed the Kitaev model on moderate size hyperbolic lattices with periodic boundary conditions. We have constructed finite quotients encoding the lattice data directly from the underlying triangle group, without passing through the construction of the non-Abelian translations of a unit cell. We have performed an exhaustive search through the compatible ones, finding that in the ground state the flux configurations can be non-uniform, because the hypotheses of Lieb’s theorem are not generally satisfied by these compact quotients \cite{Lenggenhager2025HyperbolicSpinLiquids}. While in this work we have focused solely on periodic boundaries, it would be interesting to perform a similar analysis, i.e. to identify ground state flux patterns and their structure for open boundaries as the ratio of the number of boundary and bulk lattice sites (vertices) does not vanish in the thermodynamic limit. Another interesting aspect would be to exploit the Gaussian nature of the Kitaev Hamiltonian (in a fixed flux background) to study dynamical phenomena, such as dynamics of excitations and correlations following a quench from an initial Gaussian state in analogy to studies of localization phenomena of tight-binding models on hyperbolic lattices in the presence of disorder \cite{Chen2024AndersonHyperbolic}. Another particularly natural continuation is to derive an effective Hamiltonian for the gauge-invariant flux and Wilson variables. Starting from the strongly anisotropic dimer limit, degenerate perturbation theory generates local plaquette terms together with nonlocal terms supported on non-contractible cycles. On a compact hyperbolic quotient, the plaquette length and the length of the shortest non-contractible cycle can be comparable, allowing local flux energetics to compete directly with topology-dependent contributions. This may provide an analytic explanation for the quotient-dependent nonuniform flux patterns and Wilson-sector splittings observed here. 
Finally, another direction is to treat the bond coloring itself as a physical choice. A given uncolored quotient may admit several inequivalent decompositions into perfect matchings, leading to different Hamiltonians on the same geometry. Classifying these colorings up to lattice automorphisms and comparing their flux sectors, spectra and reflection properties would disentangle effects of the quotient geometry from those of the coloring.

\acknowledgments
We would like to thank T. Bzdu\v{s}ek, P. Lenggenhager, J. Maciejko, A. Gendiar, L. Rieth, A. Safavi-Naini, K. Guo and K. Schoutens for fruitful discussions. We would also like to thank J. Ha\v{s}\'{i}k and K. Fouka for assistance at the initial stages of the project and M. Conder for helping us accessing the data we used. This work was supported by the Dutch Ministry of Economic Affairs and Climate Policy (EZK), as part of the QDNL programme.

\bibliographystyle{apsrev4-2}
\bibliography{Hyperbolic2}
\clearpage
\onecolumngrid
\appendix

\section{Physical parity projection}
\label{app:physical_parity}
We start by considering the expression of eq \ref{eq:global_parity_projector}. Expanding the product gives
\beq
P=\frac{1}{2^N}
\sum_{I\subseteq\Lambda}
\prod_{j\in I}D_j .
\eeq
For each subset $I\subseteq\Lambda$, we define
\begin{equation}
D_I:=\prod_{j\in I}D_j,
\qquad
D:=\prod_{j=1}^{N}D_j .
\end{equation}
The complementary subset $\Lambda\setminus I$ gives
\begin{equation}
D_{\Lambda\setminus I}
=
\prod_{j\notin I}D_j
=
D_I D,
\end{equation}
where we used $D_I^2=1$. Thus the full sum can be grouped into
complementary pairs $(I,\Lambda\setminus I)$, and only one representative from
each pair has to be kept:
\begin{equation}
P
=
\left(
\frac{1}{2^{N-1}}
\sum_{I}'
D_I
\right)
\left(\frac{1+D}{2}\right)
=
S\,P_0,
\end{equation}
where the prime means that only one representative from each complementary pair
is included, and
\beq
S:=
\frac{1}{2^{N-1}}
\sum_{I}'
D_I,
\qquad
P_0:=\frac{1+D}{2}.
\eeq
The operator $S$ symmetrizes over gauge-equivalent representatives, while
$P_0$ gives the global parity selection rule. In practice, once a gauge sector
is fixed, the question of whether a Majorana eigenstate survives the projection
is decided by $P_0$. We now evaluate $D$. Starting from the site-ordered product
\beq
D=\prod_{j=1}^{N}
(b_j^x b_j^y b_j^z c_j),
\eeq
we reorder the Majoranas so that all matter Majoranas $c_j$ appear together and
the remaining $b$ Majoranas are paired along bonds. We fix an orientation of
each bond, for example from the $A$ to the $B$ sublattice, and define
\beq
u_{ij}=i\,b_i^{\alpha_{ij}}b_j^{\alpha_{ij}},
\qquad
u_{ij}=\pm1.
\eeq
For a trivalent lattice, each site has coordination number $3$. Summing the
degrees over all sites gives $3N$, while each edge is counted twice, once for
each endpoint. Hence
$$
2E=3N,
\qquad
E=\frac{3N}{2},
$$
where $E$ is the number of edges. Since
\beq
u_{ij}=i\,b_i^{\alpha_{ij}}b_j^{\alpha_{ij}},
\eeq
each paired bond Majorana contributes a factor $-i$:
\beq
b_i^{\alpha_{ij}}b_j^{\alpha_{ij}}=-i\,u_{ij}.
\eeq
Therefore pairing all bond Majoranas gives
\beq
\prod_{\langle ij\rangle}
b_i^{\alpha_{ij}}b_j^{\alpha_{ij}}=(-i)^E
\prod_{\langle ij\rangle}u_{ij}.
\eeq
Using also
\beq
\pi_f:=i^{N/2}\prod_{j=1}^{N}c_j,\qquad
\prod_{j=1}^{N}c_j=i^{-N/2}\pi_f,
\eeq
the total phase accumulated in rewriting $D$ becomes
\beq
(-i)^E i^{-N/2}=(-i)^{3N/2}i^{-N/2}.
\eeq
Together with the sign coming from the Majorana reordering permutation, this
overall phase is absorbed into a geometric sign
$\eta_{\rm geom}\in\{\pm1\}$. One may therefore write

\begin{equation}
D
=
\eta_{\rm geom}\,
\pi_f
\prod_{\langle ij\rangle}u_{ij}.
\label{eq:D_eta_pi_u}
\end{equation}

$\eta_{\rm geom}$ is then fixed once the ordering conventions are
chosen. We compare
\beq
\mathcal B_{\rm site}=\bigl(b_1^x,b_1^y,b_1^z,c_1;
\dots;b_N^x,b_N^y,b_N^z,c_N\bigr)
\eeq
with
\beq
\mathcal B_{\rm edge}=
\bigl(c_1,\dots,c_N;\;b_{i_1}^{\alpha_1},b_{j_1}^{\alpha_1};\;
\dots;\;b_{i_E}^{\alpha_E},b_{j_E}^{\alpha_E}\bigr),
\eeq
where each oriented edge $(i_e,j_e)$ is taken in the same convention used to
define $u_{i_ej_e}$.

Next, diagonalizing $iA^{s}$ by an orthogonal canonical
transformation $O^{(s)}$ gives
\begin{align}
H^{s}
&=
\sum_{m=1}^{N/2}
\epsilon_m
\left(a_m^\dagger a_m-\frac12\right),\\
(-1)^{N_a}
&=
\prod_{m=1}^{N/2}
(1-2a_m^\dagger a_m).
\end{align}

To see where the determinant factor comes from, we remember that the Majorana
parity is the oriented product of all matter Majoranas. We define the initial
Majorana basis $c=(c_1,\dots,c_N)$ and the orthogonal transformation to the
canonical Majorana basis by
\beq
c_j=\sum_{k=1}^N [O^{(s)}]_{jk}\gamma_k,
\qquad O^{(s)}\in O(N).
\eeq
In the numerical implementation we extract $\det(O^{(s)})$ from the bipartite form of the Majorana matrix. Ordering the sites by sublattice, $A^s$ has the block form
\begin{equation}
A^s=
\begin{pmatrix}
0 & M^s\\
-(M^s)^{\mathsf T} & 0
\end{pmatrix}.
\end{equation}
We compute the singular-value decomposition
\begin{equation}
M^{(s)}=U^{(s)}\Lambda^{(s)}(V^{(s)})^{\mathsf T}
\end{equation}
and use
\begin{equation}
\det(O^{(s)})=\det(U^{(s)})\det(V^{(s)}).
\end{equation}
Passing from the sublattice-ordered SVD basis to the interleaved canonical Majorana basis introduces a permutation with determinant $(-1)^{n(n-1)/2}$, where $n=V/2$. For all quotients considered here, $n$ is a multiple of four, so this determinant equals $+1$.
In degenerate singular subspaces the singular-vector basis is not unique. We therefore orthonormalize the corresponding numerical basis, for instance by QR decomposition, before extracting the determinant sign. This fixes a stable representative of the same physical diagonalization and does not modify the spectrum. Because Majorana operators anticommute, the product of all $\gamma_k$ keeps
only the fully antisymmetric part of the transformation matrix. Therefore
\beq
c_1c_2\cdots c_N=\det(O^{(s)})\,\gamma_1\gamma_2\cdots\gamma_N.
\eeq
Pairing the canonical Majoranas into complex fermions,
\beq
a_m=\frac{\gamma_{2m-1}+i\gamma_{2m}}{2},\qquad m=1,\dots,N/2,
\eeq
one has
\beq
i\gamma_{2m-1}\gamma_{2m}=1-2a_m^\dagger a_m.
\eeq
Hence
\beq
i^{N/2}\gamma_1\gamma_2\cdots\gamma_N=\prod_{m=1}^{N/2}(1-2a_m^\dagger a_m)=(-1)^{N_a}.
\eeq
Combining these two identities gives
\beq
\pi_f=i^{N/2}c_1\cdots c_N=\det(O^{(s)})(-1)^{N_a}.
\eeq
Thus the matter-Majorana parity is related to the parity of the canonical
fermions by
\begin{equation}
\pi_f
=
\det\!\bigl(O^{(s)}\bigr)
(-1)^{N_a}.
\end{equation}
The global physical-state condition $D=1$ becomes
\begin{equation}
\eta_{\rm geom}\,
\det\!\bigl(O^{(s)}\bigr)\,
(-1)^{N_a}\,
\prod_{\langle ij\rangle}u_{ij}
=
1.
\label{eq:parityprojectioncondition_app}
\end{equation}

\section{Effects of projection to the physical subspace}
\label{app:projection_effects}
 We probe the effect of parity projection by comparing the lowest Majorana vacuum energy obtained before imposing the physical state constraint with the physical ground state energy $E_{\mathrm{GS}}$. The parity projection can modify the finite size ground state in two related steps. For a fixed superselection sector, the Majorana vacuum is physical only if it satisfies the required fermion parity. If it does not, the lowest physical state in that same sector is obtained by occupying the lowest energy matter mode, which raises the sector energy. Once this correction is applied independently in every sector, the ordering of the sector energies can change, so the sector with the lowest unprojected Majorana vacuum need not remain the physical ground state sector. We therefore use the difference between $E_{\mathrm{GS}}$ and the lowest unprojected vacuum energy,$\Delta_{\mathrm{parity}}=E_{\mathrm{GS}}-\min_{s}E_{\rm vac}^{(s)}$  as a direct measure of the effect introduced by the projection.
\begin{figure}[h!]
    \centering
    \includegraphics[width=0.48\linewidth]{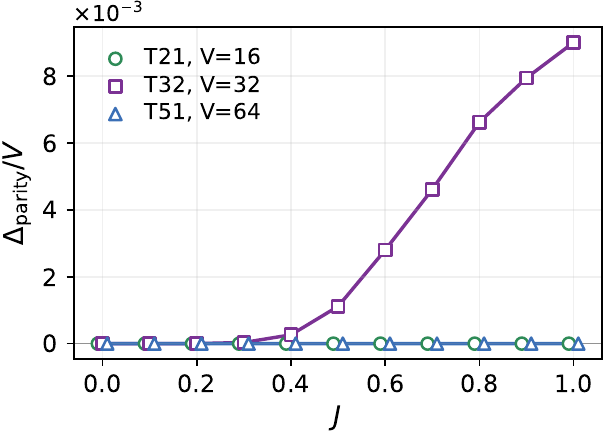}%
    \hfill
    \includegraphics[width=0.48\linewidth]{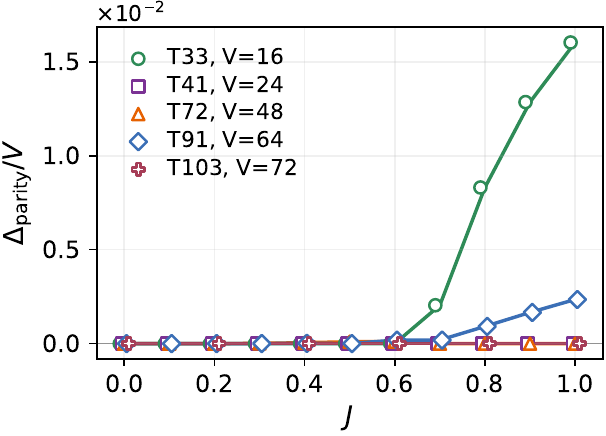}

    \smallskip
    \makebox[0.48\linewidth]{(a) $\{8,3\}$ family.}%
    \hfill
    \makebox[0.48\linewidth]{(b) $\{12,3\}$ family.}

    \caption{Parity-projection energy difference
    $\Delta_{\mathrm{parity}}/V$ as a function of $J$
    for the (a) $\{8,3\}$ and (b) $\{12,3\}$ lattice families.}
    \label{fig_DeltaParity_density}
\end{figure}

\section{Evolution of the ground state sector multiplicities with $J$}
\label{app:GS_degeneracy_J}

Table~\ref{tab:GS_degeneracy_J} reports the dependence of $d_\phi$ and $d_\omega$ along the coupling line $J_x=J_y=J$, $J_z=1$. Only the sampled values $J=0.1,0.2,\ldots,1.0$ are included. Consecutive sampled values with identical multiplicities are grouped together. The decoupled point $J=0$ is excluded because of its additional extensive many body degeneracy.

\begin{table}[h!]
\centering
\begin{tabular}{c|c|r|r|r}
\hline
Quotient & sampled $J$ & $d_\phi$ & $d_\omega$ & $d_{\mathrm{GS}}$ \\
\hline
T2.1 & $0.1$--$1.0$ & 1 & 1 & 1 \\
\hline
T3.2 & $0.1$--$0.5$ & 1 & 1 & 1 \\
     & $0.6$--$0.8$ & 4 & 1 & 4 \\
     & $0.9$        & 2 & 1 & 2 \\
     & $1.0$        & 3 & 1 & 3 \\
\hline
T3.3 & $0.1$--$0.9$ & 1 & 2 & 2 \\
     & $1.0$        & 1 & 6 & 6 \\
\hline
T4.1 & $0.1$--$0.5$ & 1 & 1 & 1 \\
     & $0.6$--$0.9$ & 2 & 1 & 2 \\
     & $1.0$        & 3 & 1 & 3 \\
\hline
T5.1 & $0.1$        & 1 & 1 & 1 \\
     & $0.2$--$0.6$ & 2 & 1 & 2 \\
     & $0.7$--$0.9$ & 1 & 1 & 1 \\
     & $1.0$        & 3 & 2 & 6 \\
\hline
T7.2 & $0.1$--$0.3$ & 1 & 2 & 2 \\
     & $0.4$--$0.9$ & 1 & 1 & 1 \\
     & $1.0$        & 1 & 3 & 3 \\
\hline
T9.1 & $0.1$--$0.9$ & 1  & 1 & 1 \\
     & $1.0$        & 48 & 4 & 192 \\
\hline
T10.3 & $0.2$--$0.8$ & 1 & 1 & 1 \\
      & $1.0$        & 3 & 1 & 3 \\
\hline
\end{tabular}
\caption{Ground state plaquette flux multiplicity $d_\phi$, Wilson multiplicity $d_\omega$, and total ground state sector degeneracy $d_{\mathrm{GS}}$ at the sampled values of $J>0$. For every entry shown, all ground state sectors belong to a single $\mathcal G_{\mathrm{phys}}(J)$ orbit.}
\label{tab:GS_degeneracy_J}
\end{table}
\end{document}